\documentclass[aps,prl,superscriptaddress,twocolumn,longbibliography]{revtex4-2}
\usepackage{amsfonts}
\usepackage{amstext}
\usepackage{amssymb}
\usepackage{amsmath}
\usepackage[utf8]{inputenc}
\usepackage{graphicx}
\usepackage{graphics}
\usepackage{cancel}
\usepackage{color,xcolor}

\usepackage{hyperref}
\hypersetup{colorlinks=true,citecolor=blue,linkcolor=blue,filecolor=blue,urlcolor=blue,pdfstartview=FitH,pdfpagemode=UseNone}
\usepackage{tikz}

\usepackage{stix}
\definecolor{boxcolor}{HTML}{108f64}

\bibpunct{[}{]}{,}{n}{}{}

\begin{document}

\title{Quantum geometry and the electric magnetochiral anisotropy in noncentrosymmetric polar media}

\author{Pierpaolo Fontana}
\affiliation{Departament de Física, Universitat Autònoma de Barcelona, 08193 Bellaterra, Spain}

\author{Victor Velasco}
\affiliation{International School for Advanced Studies (SISSA), Via Bonomea 265, I-34136 Trieste, Italy}

\author{Chang Niu}
\affiliation{Elmore Family School of Electrical and Computer Engineering, Purdue University, West Lafayette, Indiana 47907, United States} 
\affiliation{Birck Nanotechnology Center, Purdue University, West Lafayette, Indiana 47907, United States}

\author{Peide D. Ye}
\affiliation{Elmore Family School of Electrical and Computer Engineering, Purdue University, West Lafayette, Indiana 47907, United States} 
\affiliation{Birck Nanotechnology Center, Purdue University, West Lafayette, Indiana 47907, United States}

\author{Pedro V. Lopes}
\affiliation{Instituto de F\'isica, Universidade Federal do Rio de Janeiro, Caixa Postal 68528, 21941-972 Rio de Janeiro, Brazil}

\author{Kaio E. M. de Souza}
\affiliation{Universidade Federal do Rio de Janeiro - Campus UFRJ Duque de Caxias, Rod. Washington Luiz, 19593 - km 104,5 - Santa Cruz da Serra, Duque de Caxias, 25240-005, Rio de Janeiro, Brazil}

\author{Marcus V. O. Moutinho}
\affiliation{Universidade Federal do Rio de Janeiro - Campus UFRJ Duque de Caxias, Rod. Washington Luiz, 19593 - km 104,5 - Santa Cruz da Serra, Duque de Caxias, 25240-005, Rio de Janeiro, Brazil}

\author{Caio Lewenkopf}
\affiliation{Instituto de F\'isica, Universidade Federal do Rio de Janeiro, Caixa Postal 68528, 21941-972 Rio de Janeiro, Brazil}

\author{Marcello B. Silva Neto}
\affiliation{Instituto de F\'isica, Universidade Federal do Rio de Janeiro, Caixa Postal 68528, 21941-972 Rio de Janeiro, Brazil}

\date{\today}

\begin{abstract}

The electric magnetochiral anisotropy is a nonreciprocal phenomenon accessible via second harmonic transport in noncentrosymmetric, time-reversal invariant materials, in which the rectification of current, ${\bf I}$, can be controlled by an external magnetic field, ${\bf B}$. Quantum geometry, which characterizes the topology of Bloch electrons in a Hilbert space, provides a powerful description of the nonlinear dynamics in topological materials. Here, we demonstrate that the electric magnetochiral anisotropy in noncentrosymmetric polar media owes its existence to the quantum metric, arising from the spin-orbit coupling, and to large Born effective charges. In this context, the reciprocal magnetoresistance $\beta{\bf B}^2$ is modified to $R( I,P,B)=R_0[1+\beta B^2 + \gamma^{\pm}{\bf I}\cdot({\bf P}\times{\bf B})]$, where the chirality dependent $\gamma^{\pm}$ is determined by the quantum metric dipole and the polarization ${\bf P}$. We predict a universal scaling $\gamma^{\pm}(V)\sim V^{-5/2}$ which we verified by phase sensitive, second harmonic transport measurements on hydrothermally grown 2D tellurium films under applied gate voltage, $V$. The control of rectification by varying ${\bf I}$, ${\bf P}$, ${\bf B}$, and $V$, demonstrated in this work, opens up new avenues for the building of ultra-scaled CMOS circuits.

\end{abstract}

\maketitle

{\it Introduction.--} 
Nonreciprocal transport phenomena occur when the flow of charge, such as electrons and holes in semiconductors, depends on the direction of the current, resulting in an asymmetric conduction. In $pn$ junctions, this nonreciprocity arises from the interface between $p$-type and $n$-type semiconductor materials, where a built-in electric field creates a depletion region
that allows current to flow more easily in one direction than in the other \cite{Fundamental-Electric-Circuits}. This behavior is fundamental to the operation of diodes, acting as one-way gates for current flow and enabling rectification of alternating currents, signal modulation, and voltage regulation. These features highlight the indispensable role of nonreciprocal transport phenomena in modern technology \cite{Fundamental-Power-Electronics}.

One of the basic ingredients of nonreciprocal phenomena is the lack of inversion symmetry in chiral molecules, films, or crystals \cite{ReviewTokuraNagaosa}. 
Noncentrosymmetry leads to directional propagation of quantum particles, like photons, spins, phonons, and electrons, giving rise to various phenomena. These include the natural optical activity in chiral materials \cite{NaturalOpticalActivity}, the nonreciprocal magnon transport or spin current in chiral magnets \cite{Chiral-Magnons}, the conversion of a coherent lattice vibration into a quasi-static structural distortion in nonlinear phononics \cite{PhononicRectification}, and the unidirectional magnetoresistance, or electrical magnetochiral anisotropy (eMChA), in polar and chiral semiconductors \cite{eMChA-Polar-Semicond}. The latter, eMChA, is a particular type of nonreciprocal phenomenon observed in materials with a large magnetoresistance, such as a Si-based field effect transistor (FET) \cite{Rikken_2005}. Other systems were also found to exhibit nonreciprocal magnetoresistance, such as the polar semiconductor BiTeBr \cite{PolarSemiconductor}, the multiferroic semiconductor (Ge,Mn)Te \cite{MultiferroicGeMnTe}, the topological insulator nanowire heterostructure (Bi$_{1-x}$Sb$_2$)$_2$Te$_3$ under an external voltage bias \cite{SurfaceTINanowireHeterostructure}, twisted bilayer graphene \cite{eChMAinTBG}, and the topological semimetal ZrTe$_5$ \cite{eMChAinZrTe5}. In all these systems, the ability to fully control rectification using external parameters such as current, polarization, and magnetic field opens new possibilities for the development of, for instance, ultra-scaled complementary metal–oxide–semiconductor (CMOS) circuits \cite{Dalvin}.

\begin{figure}[t]
\includegraphics[scale=0.34]{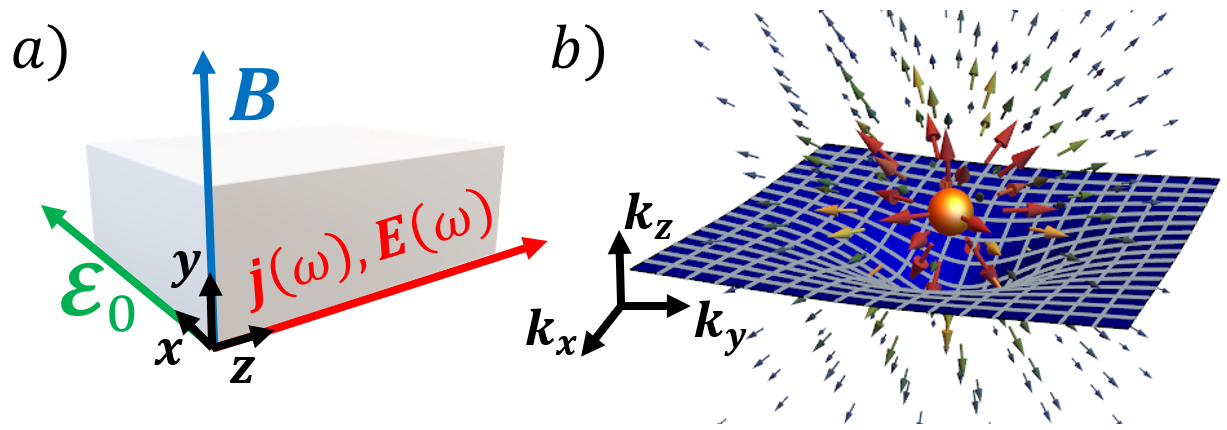}
\caption{(a) Set up used in this work. The static and uniform electric, $\boldsymbol{\cal E}_0\parallel x$, and magnetic, ${\bf B}\parallel y$, fields are perpendicular to the flow of the ac-current ${\bf j}(\omega)\parallel z$ and voltage related, ac-electric field ${\bf E}(\omega)\parallel z$. (b) Topological singularities, such as a Wey node, introduce both a radial spin texture (and Berry curvature) as well as momentum space geodesics (white grid) for the adiabatic evolution of quantum states.}
\label{fig:experimental.setup}
\end{figure}

The underlying mechanism responsible for the nonreciprocal carrier diffusion can be elegantly understood through a heuristic argument first proposed by Rikken \cite{Rikken_2005}. This mechanism is the transport analog of an optical phenomenon observed when light with wave vector ${\bf k}$ propagates perpendicularly to the crossed electric (${\bf E}$) and magnetic (${\bf B}$) fields \cite{Rikken_2002}. In this context, the effective refractive index receives a relativistic, nonreciprocal contribution of the form $\delta n\sim{\bf k}\cdot{\bf v}$, obtained by means of a Lorentz transformation to a moving frame with velocity ${\bf v}=c{\bf E}\times{\bf B}/B^2$ with respect to the laboratory  \cite{jackson_classical_1999}. Rikken argued that an analogous mechanism applies to diffusive transport \cite{Rikken_2005}. When a current ${\bf I}\sim\langle{\bf k}\rangle$ flows perpendicularly to crossed electric ($\boldsymbol{\cal E}_0$) and magnetic (${\bf B}$) fields, see Fig. \ref{fig:experimental.setup}(a), the reciprocal magnetoresistance, $\Delta R/R=1+\beta{\bf B}^2$, acquires a nonreciprocal, relativistic correction of the form $\delta R/R=2\gamma {\bf I}\cdot(\boldsymbol{\cal E}_0\times{\bf B})$. This correction arises because, in the laboratory frame, the motion of a charged particle acquires an {\it additional drift velocity} ${\bf v}=c\boldsymbol{\cal E}_0\times{\bf B}/B^2$. 
The field $\boldsymbol{\mathcal{E}}_0$ can represent an applied electric field, the field arising from the net polarization ${\bf P}$ in noncentrosymmetric materials, or the field resulting from a band structure offset at the interface of semiconductor heterostructures \cite{Rikken_2005}. Furthermore, Rikken conjectured that the nonlinear conductivity should have the form $\tilde{\sigma}_{ij}(\langle{\bf k}\rangle\cdot\boldsymbol{\cal E}_0\times{\bf B})$ \cite{Rikken_2005}, extending the Onsager´s reciprocity relations for time-reversal invariant systems \cite{HertelLecturesOT}, and it would be accessible via second harmonic transport experiments.

In this Letter, we demonstrate that Rikken´s conjecture follows naturally from the quantum geometric properties \cite{QGinCondMattReview} of noncentrosymmetric media, which are characterized by strong spin-orbit interactions (relativistic corrections) and large Born effective charges (macroscopic polarization). Using Boltzmann's semiclassical approach to the electronic transport \cite{Ziman}, we show that including the quantum metric dipole correction to the carriers velocity \cite{Kaplan_2024_PRL,KaplanNature,KaplanPRR,Gravity} in the Lorentz force results in a relativistic correction to the nonlinear magnetoconductivity of the form $\tilde{\sigma}_{ij}(\langle{\bf v}\rangle\cdot\boldsymbol{\cal E}_0\times{\bf B})$. Here, $\langle{\bf v}\rangle$ is the average velocity weighed by the quantum metric dipole and $\boldsymbol{\cal E}_0$ denotes the electric field associated to the net polarization of the system. In order to verify our results we performed density functional theory (DFT) calculations to estimate the polar electric field, as well as phase sensitive, second harmonic transport measurements in 2D $n-$type chiral tellurium films at low temperature, varying both the applied magnetic field and the gate voltage. 

{\it Second harmonics.--} 
The polarization-induced eMChA is a nonlinear phenomenon described by the relation 
\begin{equation}
R(I,B,P)=R_0[1+\beta B^2 + \gamma^{\pm}{\bf I}\cdot({\bf P}\times{\bf B})],   
\label{eMChA}
\end{equation}
where $R_0$ is the resistance, $\beta$ is the reciprocal magnetoresistance coefficient, and $\gamma^\pm$ is the nonreciprocal eMChA coefficient for the two possible chiralities \cite{Rikken_2019,Rikken_2002,Rikken_2005}. This phenomenon is described by a fourth-rank tensor, ${\bf G}$, in the expansion of the AC current density, ${\bf j}$, in powers of the AC electric field strength and the external magnetic field \cite{Spivak_2023}
\begin{equation}
    j_i=\sigma_{ij} E_j+\sigma^{(H)}_{ijk}E_j B_k+G_{ijk\ell}E_j E_k B_\ell.
    \label{Nonlinear-Current}
\end{equation}
Here, $\sigma_{ij}$ is the linear conductivity, and $\sigma^{(H)}_{ijk}$ represents the Hall conductivity. The eMChA effect is encoded in the tensor ${\bf G}$, symmetric in $j$ and $k$. By solving for ${\bf E}$ as a function of ${\bf B}$ and ${\bf j}$ up to order $Bj^2$, the tensor ${\bf G}$ is \cite{Spivak_2023}
\begin{equation}
    G_{ijk\ell}\propto \gamma_{ij^\prime k^\prime \ell}\sigma_{j^\prime j}\sigma_{k^\prime k},
    \label{eq.G-to-Gamma}
\end{equation}
where the fourth-rank tensor $\gamma_{ijk\ell}$ characterizes the second-harmonic generation described by \cite{Rikken_2019,Spivak_2023}
\begin{equation}
    E_i^{2\omega}=\gamma_{ijkl}j_j^\omega j_k^\omega B_l,
\end{equation}
under conditions of a very long period $T = 2\pi/\omega$. The normalized nonreciprocal resistance therefore corresponds to \cite{supplmat}
\begin{equation}
    \frac{4 V_{zz}^{2\omega}}{V_{zz}^\omega}=\frac{\Delta R}{R_0}=
    2\gamma^{\pm}{\bf I}\cdot({\bf P}\times{\bf B}), 
    \label{eq.polar.emcha.coeff}
\end{equation}
and the normalized eMChA coefficient above, $\gamma^{\pm}$, introduced in Eq. \eqref{eMChA} for an isotropic system, scales as $\gamma^\pm\sim G/\sigma$.

{\it Quantum Geometry.--} 
The existence of a nonlinear conductivity, even at the smallest perturbation, has been recently claimed to have a quantum geometric origin \cite{QGinCondMattReview}. Quantum mechanics can be formulated as a geometric theory in a Hilbert space, where the distance between two adjacent quantum states, $\left|n_{\bf k}\right.\rangle$ and $\left|n_{{\bf k}+d{\bf k}}\right.\rangle$, in the $n-$th band is fully determined by the {\it quantum geometric tensor} \cite{TopandGeoAspectsBandTheory}
\begin{equation}
    {\cal \bf Q}^n({\bf k})={\bf g}^n({\bf k})-\frac{i}{2}{\bf \Omega}^n({\bf k}).
\end{equation}
Its imaginary part, the Berry curvature, ${\bf \Omega}^n({\bf k})$, has long been known to generate several important linear transport phenomena, such as the quantum anomalous Hall effect \cite{XiaoRMP,Hedgehog}, since it plays the role of a magnetic field in reciprocal space. On the other hand, its real part, the quantum metric, ${\bf g}^n({\bf k})$, has only recently been addressed experimentally, both by probing nonlinear transport quantities, such as the nonlinear Hall effect and nonlinear magnetoresistance \cite{KaplanNature}, and via angle-resolved photoemission spectroscopy \cite{Kang2024QGT}. The quantum metric imposes a kinematic constraint for Bloch electrons to follow momentum space geodesics \cite{Gravity}, thereby modifying the wave packet dynamics, as in Fig. \ref{fig:experimental.setup}(b). The calculation of ${\bf g}^n({\bf k})$ and ${\bf \Omega}^n({\bf k})$ is straightforward for a $2\times 2$ two-band Hamiltonian ${\cal H}={\bf d}(\bf k)\cdot\sigma$. In this case, $n=\pm$ and, with $\partial_a\equiv\partial/\partial k_a$, we have
\begin{eqnarray}
    {\bf g}^\pm_{ab}&=&
    \frac{1}{4 d^2}\left[\partial_a{\bf d}\cdot\partial_b{\bf d}-\frac{1}{d^2}(\partial_a{\bf d}\cdot{\bf d})(\partial_b{\bf d}\cdot{\bf d})\right],\nonumber\\
    {\bf \Omega}^\pm_{ab}&=&\mp\frac{(\partial_a{\bf d}\times\partial_b{\bf d})\cdot{\bf d}}{2 d^3}.
    \label{QM-and-BC}
\end{eqnarray}
%

{\it Boltzmann transport.--} 
We compute the current density
\begin{equation}
    {\bf j}=-e\sum_{n,{\bf k}}\dot{\bf r}_n\,f_n({\bf k}),
\end{equation}
up to order $E^2 B$, as shown in Eq. (\ref{Nonlinear-Current}). Here, $f_n({\bf k})$ is the non-equilibrium distribution function, satisfying the Boltzmann transport equation in the relaxation time approximation \cite{Ziman}
\begin{equation}
        \frac{\partial f_n({\bf k})}{\partial t}+\dot{\bf r}_n\cdot\frac{\partial f_n({\bf k})}{\partial {\bf r}}+\dot{\bf k}_n\cdot\frac{\partial f_n({\bf k})}{\partial {\bf k}}=-\frac{f_n({\bf k})-f_n^0({\bf k})}{\tau_{\bf k}},
\end{equation}   
where $f_n^0({\bf k})=1/(e^{\beta(\varepsilon_n({\bf k})-\mu)}+1)$ is the equilibrium Fermi-Dirac distribution, $\varepsilon_n({\bf k})$ is the dispersion relation of the $n-$band, $\beta=1/k_B T$ is the inverse temperature, $\mu$ is the chemical potential, and $\tau_{\bf k}=\tau$ is the relaxation time. We consider an homogeneous system, \textit{i.e.}, $\partial f_n/\partial {\bf r}=0$. The velocity, $\dot{\bf r}_n$, and the acceleration, $\dot{\bf k}_n$, are defined by the semiclassical equations of motion \cite{Gravity}
\begin{eqnarray}
    \dot{\bf r}_n&=&{\bf v}^n_{\bf k}+
    \dot{\bf k}_n\times{\bf \Omega}^n_{\bf k}+\dot{\bf k}_n\cdot{\bf \Gamma}^n(\partial{\bf g})\cdot\dot{\bf k}_n,\label{Velocity}\\
    \hbar\dot{\bf k}_n&=&-e({\bf E}+{\boldsymbol{\cal E}}_0)-e\dot{\bf r}_n\times{\bf B},
    \label{Acceleration}
\end{eqnarray}
where ${\boldsymbol{\cal E}}_0$ is a static and uniform electric field resulting from the macroscopic polarization and we assumed $c=1$. In the equation for $\dot{\bf r}_n$, the first term, ${\bf v}^n_{\bf k}=(1/\hbar)\nabla_{\bf k}\varepsilon_n({\bf k})$, represents the band velocity; the second term, $\dot{\bf k}_n\times{\bf \Omega}^n_{\bf k}$, is the anomalous velocity, arising from the Berry curvature; the third term is the geodesic contribution due to the quantum metric dipole, with $\dot{\bf k}_n\cdot{\bf \Gamma}^n(\partial{\bf g})\cdot\dot{\bf k}_n\equiv{\bf \Gamma}^n_{ij\ell}\dot{\bf k}_{n,j}\dot{\bf k}_{n,\ell}$, represented by the Christoffel symbols ${\bf \Gamma}^\pm_{ij\ell}=2\partial_{i}\tilde{\bf g}^\pm_{j\ell}-(\partial_{j}\tilde{\bf g}^\pm_{\ell i}+\partial_{\ell}\tilde{\bf g}^\pm_{ij})/2$, in terms of the energy normalized quantum metric, $\tilde{\bf g}^\pm_{ij}={\bf g}^\pm_{ij}/2(\varepsilon_\pm-\varepsilon_\mp)$ \cite{Kaplan_2024_PRL}. 

In the DC limit and in the presence of a static and uniform magnetic field ${\bf B}$, the non-equilibrium distribution function $f_n({\bf k})=f_n^0({\bf k})+\delta f_n({\bf k})$ has a correction due to the Lorentz force \cite{supplmat}, as in Eq. \eqref{Acceleration}
\begin{equation}
    \small
    \delta f_n({\bf k})=e\tau\left\{(\dot{\bf r}_n\cdot{\bf E})+e\tau\left(\frac{\dot{\bf r}_n\times{\bf B}}{\hbar}\right)\cdot\nabla_{\bf k}(\dot{\bf r}_n\cdot{\bf E})\right\}\left(\frac{\partial f_0}{\partial\varepsilon}\right).\nonumber
\end{equation}
We then substitute the expression for $\dot{\bf r}_n$ from Eq. \eqref{Velocity} into $\delta f_n({\bf k})$, and we select the contribution due to the quantum metric of the form $2\,e^2\,{\bf E}\cdot{\bf \Gamma}(\partial{\bf g})\cdot{\boldsymbol{\cal E}}_0/\hbar^2$. The nonreciprocal contribution to the total current at order $E^2 B$ becomes 
\begin{equation}
\delta{\bf j}=\frac{2e^5\tau^2}{\hbar^3}\sum_{n,{\bf k}}{\bf v}^n_{\bf k}
[({\bf E}\cdot{\bf \Gamma}^n(\partial{\bf g})\cdot\boldsymbol{\cal E}_0)\times{\bf B}]
\cdot\nabla_{\bf k}({\bf v}^n_{\bf k}\cdot{\bf E})
    \left(-\frac{\partial f_0}{\partial \varepsilon_n}\right),\nonumber
\end{equation}
This expression explicitly contains the structure $\langle{\bf v}^n_{\bf k}\rangle\cdot({\boldsymbol{\cal E}}_0\times{\bf B})$, validating Rikken's conjecture \cite{Rikken_2005}. The momentum dependent Christoffel symbols, $\boldsymbol{\Gamma}({\bf k})$, emerge as a quantum metric dipole contribution to the nonlinear conductivity. Writing the above expression in tensor form as $\delta j_i=G_{ijk\ell}E_j E_k B_\ell$, we conclude that the eMChA tensor is given by the expression
\begin{equation}
    G_{ijk\ell}=\frac{2e^5\tau^2}{\hbar^3}
    \sum_{n,{\bf k}}{v}^n_{{\bf k},i}
    \left\{\epsilon_{ab\ell}{\Gamma}^n_{bkm}(\partial{\bf g})\boldsymbol{\cal E}_{0,m}\right\}\,\partial_a {v}^n_{{\bf k},j}
    \left(-\frac{\partial f_0}{\partial \varepsilon_n}\right).
    \label{G-tensor}
\end{equation}
In the above expression, the Levi-Civita symbol is defined as $\epsilon_{xyz}=+1$, and, at low temperatures, the equilibrium distribution derivatives reduces to $(-\partial f_n^0/\partial\varepsilon_n)=\delta(\varepsilon_n({\bf k})-\mu)$. The $G_{ijk\ell}$ tensor is nonzero because both ${\bf v}_{\bf k}$ and ${\bf \Gamma}({\bf k})$ are odd under the transformation ${\bf k}\rightarrow -{\bf k}$, and in the presence of a macroscopic polarization the electric field $\boldsymbol{\cal E}_0\neq 0$. Most importantly,  $G_{ijk\ell}\sim\tau^2$ and since from Eq. \eqref{eq.G-to-Gamma} $G_{ijk\ell}\sim\gamma_{ijk\ell}\sigma^2$, with $\sigma\sim\tau$, we conclude that the eMChA tensor, $\gamma_{ijk\ell}$, is intrinsic and does not depend on extrinsic mechanisms, through $\tau$, being determined entirely by the quantum geometric properties of the material.

\begin{figure}[t]
\includegraphics[scale=0.32]{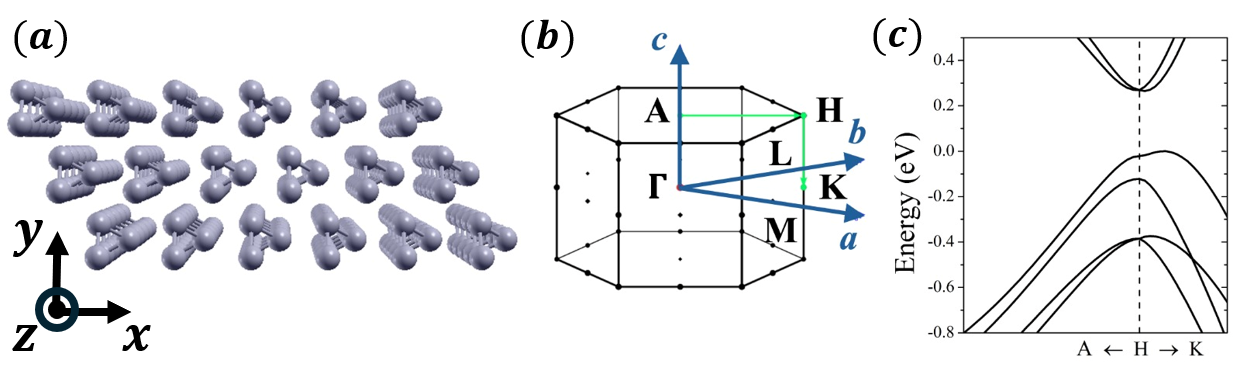}
\caption{(a) The trigonal $\alpha-$Te crystal structure consists of spiral chains with three-fold screw symmetry held together by inter-chain van der Waals bonds. (b) First Brillouin zone of Te. (c) band features near the direct band edge at the H point along $A-H-K$ directions.}
\label{fig:crystal-and-band-structures}
\end{figure}

{\it $n-$type 2D tellurium.--} 
Tellurium (Te) is a narrow-gap Weyl semiconductor composed of one-dimensional chiral chains arranged into a hexagonal lattice due to van der Waals interactions, and its $\alpha-$Te form is shown in Fig. \ref{fig:crystal-and-band-structures}(a). Its two-dimensional version has attracted significant scientific and technological interest due to its promising electronic, optoelectronic, and piezoelectric applications. \cite{Resurrection_2022}. The reason for studying $n-$type 2D tellurium is twofold: first, $n-$type tellurium hosts a Weyl node at the $H-$point in the Brillouin zone, close to the minimum of the conduction band, see Figs. \ref{fig:crystal-and-band-structures}(b) and (c), which ensures strong quantum geometric properties, characterized by ${\bf \Gamma}_{ijk}\neq 0$; second, 2D tellurium films exhibit a nonzero net polarization \cite{Ferro-and-Piezo-Tellurium}, arising from the large Born effective charges associated with the lone pairs, resulting in ${\boldsymbol{\cal E}}_0\neq 0$. Furthermore, in $\alpha-$Te films with thicknesses of only a few nanometers, the population in the conduction band can be controlled by top and back gating \cite{Chang_2023}, 
enabling the investigation of the $G-$tensor in Eq. \eqref{G-tensor} as a function of the chemical potential, $\mu$.    

\begin{center}
\begin{table}
\begin{tabular}{|p{1.3cm}|p{1.3cm}|p{1.3cm}|p{1.3cm}|p{1.3cm}|p{1.3cm}|}
 \hline
 \multicolumn{6}{|c|}{Conduction Band parameters in ${\cal H}_c$} \\
 \hline
 A $\times 10^{-15}$ & B $\times 10^{-15}$ & S $\times 10^{-9}$ & C $\times 10^{-9}$ & F $\times 10^{-15}$ & G $\times 10^{-21}$ \\
  ($eV cm^{2}$) & ($eV cm^{2}$) & ($eV cm$) & ($eV cm$) & ($eV cm^{2}$) & ($eV cm^{2}$) \\
 \hline
 $6.7$   & $4.2$ & $5.8$ & $3.6$ & $1.7$ &  $0.2$ \\
 \hline
\end{tabular}
\caption{Table I: experimental values for the parameters of Hamiltonian (\ref{Hamiltonian-Conduction-Band}) obtained from magneto-optical transition \cite{Magnetooptical}.}
\label{Table-I}
\end{table}
\end{center}

The first two ingredients necessary to proceed are the components of the velocity, ${\bf v}_{\bf k}$, and of the Christoffel symbols, ${\bf \Gamma}({\bf k})$, appearing in Eq. \eqref{G-tensor}. The conduction band structure of tellurium has been derived using ${\bf k}\cdot{\bf p}$ perturbation theory, yielding the $2\times 2$ Hamiltonian \cite{Magnetooptical}
\begin{equation}
{\cal H}_c=
\begin{bmatrix}
A k_z^2 + B k_\perp^2 + S k_z & C k_- + iG k_- k_z + F k_+^2 \\
C k_+ - iG k_+ k_z + F k_-^2 & A k_z^2 + B k_\perp^2 - S k_z
\end{bmatrix}.
\label{Hamiltonian-Conduction-Band}
\end{equation}
Up to second order in ${\bf k}\cdot{\bf p}$, the Hamiltonian includes the $S k_z$ term, a linear-in-$k_z$ contribution due to the lack of inversion symmetry, and the parabolic $A k_z^2 + B k_\perp^2$ contributions. The combination of ${\bf k}\cdot{\bf p}$ and the spin-orbit interaction gives rise to the off-diagonal, linear-in-$k_\pm$ band mixing elements $C k_\pm$, generating a principal Weyl node at the $H-$point. We also include a trigonal warping term $F k_\pm^2$, responsible for the generation of satellite Weyl nodes away from the $H-$point, and a term arising from the ${\bf k}-$dependent spin-orbit interaction, \textit{i.e.}, $\pm iG k_\mp k_z$. All the parameters in the Hamiltonian \eqref{Hamiltonian-Conduction-Band} can be obtained experimentally, {\it e.g.}, through magneto-optical transitions \cite{Magnetooptical}, with values listed in Table \ref{Table-I}.

The $2\times 2$ Hamiltonian \eqref{Hamiltonian-Conduction-Band} can be expanded on the basis of the identity and Pauli matrices as ${\cal H}=d_0({\bf k})\mathcal{I}+{\bf d}(\bf k)\cdot\sigma$, with
\begin{eqnarray}
    d_0({\bf k})&=&A k_z^2 + B k_\perp^2,\nonumber\\
    d_x({\bf k})&=&C k_x + G k_y k_z + F(k_x^2 - k_y^2),\nonumber\\
    d_y({\bf k})&=&C k_y - G k_x - 2 F k_x k_y,\nonumber\\
    d_z({\bf k})&=&S k_z.
 \end{eqnarray}
Both the Berry curvature, ${\bf \Omega}^n_{\bf k}$, and the quantum metric, ${\bf g}^n_{\bf k}$, can now be computed using Eqs. \eqref{QM-and-BC}. The $G_{zzzy}$ component of the eMChA tensor in Eq. \eqref{G-tensor}, can be evaluated numerically, as discussed in the Supplemental Material (SM) \cite{supplmat}. To simplify the analysis and gain physical insight, we neglect both the trigonal warping, $F=0$, and the ${\bf k}-$dependent spin-orbit interaction, $G=0$, which are the smallest parameters in Table \ref{Table-I}. Within these assumptions, the conduction bands are described by
\begin{equation}
    \varepsilon_\pm({\bf k})=A k_z^2 + B k_\perp^2 \pm \sqrt{S^2 k_z^2+C^2 k_\perp^2}.
    \label{Two-Bands}
\end{equation}
The $z-$component of the velocity, $v^\pm_z({\bf k})$, entering in Eq. \eqref{G-tensor}, for these bands and the $xzx$ component of the Christoffel symbol, $\Gamma_{xzx}({\bf k})$, can also be found analytically as
\begin{eqnarray}
    v^\pm_z({\bf k})&=&2A k_z\pm\frac{k_z S^2}{\sqrt{S^2 k_z^2+C^2 k_\perp^2}}\nonumber\\
    \Gamma^\pm_{xzx}({\bf k})&=&\pm\frac{5 S^2 C^4 k_x^2 k_z}{8(C^2 k_\perp^2+S^2k_z^2)^{7/2}}.
\end{eqnarray}
Away from the Weyl node we can use a parabolic approximation, $\varepsilon_\pm({\bf k})\approx \hbar^2{\bf k}^2/2m_\pm^*$. In this case, the velocity and its $k_z-$derivative are simply given by, ${\bf v}^\pm_{\bf k}=\hbar{\bf k}/m^*_\pm$ and $\partial_z{v}^\pm_{{\bf k},i}=\delta_{z,i}\,\hbar/m^*_\pm$. The scaling of the normalized eMChA coefficient appearing in Eq. \eqref{eq.polar.emcha.coeff}, with the chemical potential, $\mu$, can now be obtained exactly, $\gamma^\pm(\mu)\equiv G_{zzzy}(\mu)/\sigma(\mu)$, with the scaling of the $G_{zzzy}(\mu)$ tensor itself given by \cite{supplmat}
\begin{equation}
    G_{zzzy}(\mu)=\frac{G_0}{\mu^{3/2}},
    \label{eq.universal.scaling}
\end{equation}
where $G_0={\cal V} e^5\tau^2\boldsymbol{\cal E}_{0,x}{\cal I}((m_+^*)^{-5/2}-(m_-^*)^{-5/2})/(2\pi)^2\sqrt{2}$, while ${\cal V}$ is the volume of the 2D Brillouin zone and 
\begin{equation}
    {\cal I}=\int_0^{2\pi}d\alpha 
    \frac{5S^2 C^4\sin^2(\alpha)\; \cos^2(\alpha)}{8(C^2 \sin^2(\alpha)+S^2 \cos^2(\alpha))^{7/2}}.
\end{equation}
%

\begin{figure}[!h]
    \centering
    \includegraphics[scale=0.4]{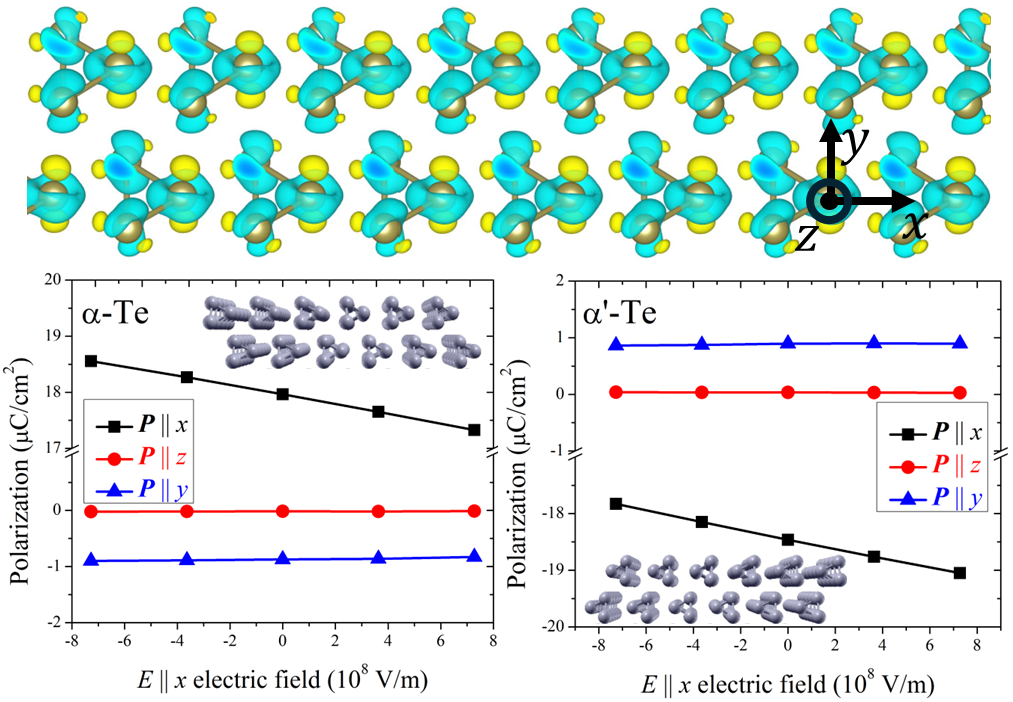}
    \caption{The large Born effective charges due to lone pairs give rise to a net polarization, ${\bf P}\parallel\hat{\bf x}$, in 2D thin films of tellurium, that is perpendicular to both the $\hat{\bf z}$ (helices) and $\hat{\bf y}$ (growth) directions. The upper pannel shows the charge distribution in $\alpha-$Te, with the excess charge represented in yellow and charge deficit in blue. The macroscopic polarization is also shown in both  bottom pannels for $\alpha$ and $\alpha^\prime$ Te, as a function of an applied electric field oriented along the $x-$direction. }
    \label{fig:polarization-lone-pairs}
\end{figure}

The final ingredient in Eq. \eqref{G-tensor}, the electric field $\boldsymbol{{\cal E}}_0$, arises from the macroscopic polarization in the 2D tellurium film. The geometry depicted in Fig. \ref{fig:experimental.setup} requires that this polarization is confined to the plane of the film and perpendicular to both the direction of the helices (${\bf z}$) and the growth direction (${\bf y}$), so that ${\bf P}\parallel{\bf x}$. In order to verify if this requirement is satisfied, we performed DFT calculations using the Quantum Espresso \cite{Giannozi2009} and Siesta \cite{Soler_2002} packages for a two-layer Te film. These computations allowed us to determine the charge density and macroscopic polarization as a function of an applied external electric field across all possible directions. The computational details are presented in the SM \cite{supplmat} and the results are presented in Fig. \ref{fig:polarization-lone-pairs}. Our results show that the polarization is indeed oriented along the $x-$axis, with a  magnitude of roughly $\pm 18\, \mu C/cm^2$ at zero applied electric field, for $\alpha-$Te ($+$) and $\alpha^\prime-$Te ($-$) \cite{Ferro-and-Piezo-Tellurium}. We have also observed a small polarization component perpendicular to the film, while the component along the coil direction was nearly zero. The polarization is robust under variations of an external field applied along the $x-$axis, with its $x$-component either increasing or decreasing based on the orientation of the field. This provides another way of rectification control, potentially of interest for electronic applications.

{\it Discussion.--} 
We now compare our results to second harmonic transport experiments in 2D $\alpha-$Te with the setup described in Fig. \ref{fig:experimental.setup}(a) \cite{Chang_2023}. Phase sensitive measurements allows access to the resistance difference between the currents flowing in $+I$ and $-I$ directions, {\it i.e.}, $\Delta R\equiv R(B,I)-R(B,-I)$. A driving AC current, $I^\omega$, is typically injected into the system and a longitudinal voltage $V_{zz}^\omega$, as well as its second harmonic, $V_{zz}^{2\omega}$, are measured \cite{Chang_2023}. 
The resulting eMChA coefficient satisfies Eq. (\ref{eq.polar.emcha.coeff}) and is linear in both ${\bf B}$ and ${\bf I}$, as reported in \cite{Chang_2023}. Furthermore, for scan-rotations of ${\bf B}$ confined to either the $y-z$ ($\theta$ rotation) or $x-y$ ($\phi$ rotation) planes, the angular dependence of the eMChA coefficient follows cosine functions of $\theta$ and $\phi$, respectively, again satisfying Eq. (\ref{eq.polar.emcha.coeff}) \cite{supplmat} for the largest polarization component, ${\bf P}\parallel {\bf x}$ . Finally, the small (but nonzero) component of ${\bf P}\parallel -{\bf y}$, perpendicular to the film, accounts for the much smaller eMChA coefficient $(\sim 10^{-6} T^{-1})$, for the scan-rotation of ${\bf B}$ confined to the $z-x$ plane ($\alpha$ rotation) \cite{supplmat}. 

\begin{figure}[t]
    \includegraphics[scale=0.43]{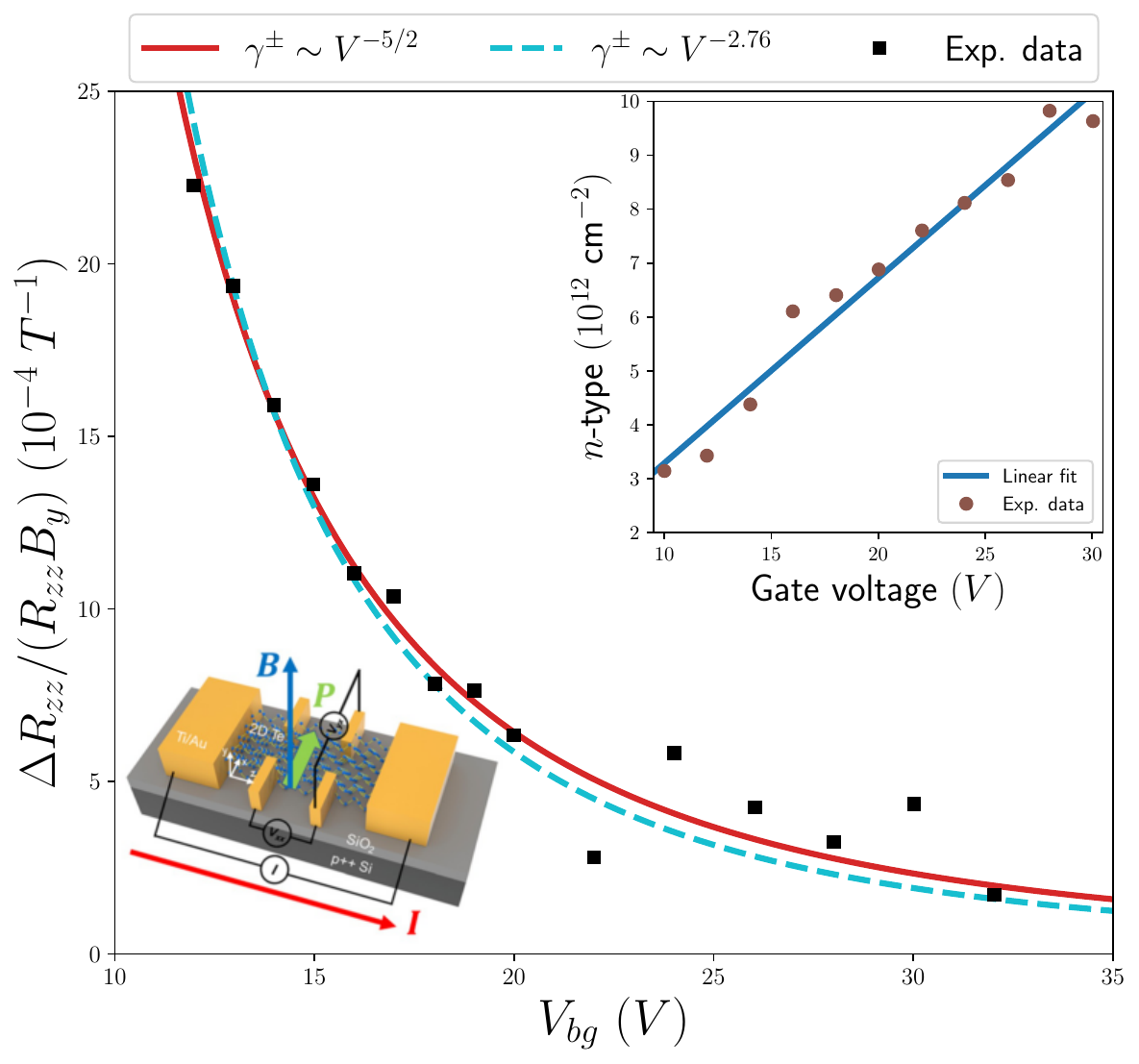}
    \caption{Experimental data of $\Delta R_{zz}/(R_{zz}B_y)$, from \eqref{eq.polar.emcha.coeff}, as a function of the backgate voltage, demonstrating the $\gamma^\pm(V)\sim V^{-5/2}$ scaling (red curve). We also superimpose to the experimental data the $\gamma^\pm(V)\sim V^{-2.76}$ (dashed cyan curve), whose exponent has been obtained through the numerical computation of $G_{zzzy}$ for the full conduction bands. The inset shows the carrier concentration, $n$, as a function of gate voltage, $V$, for 2D tellurium. (inset) The $20$nm tellurium flakes were transferred onto $90$nm SiO$_2$/Si substrate.}
    \label{fig:rectification.angle}
\end{figure}

In order to test the universal scaling predicted in Eq. (\ref{eq.universal.scaling}), a gate voltage was used to control the population of the conduction band with $n-$type carriers \cite{Chang_2023}, through $n(V)=\kappa V$, see Fig. \ref{fig:rectification.angle} (inset). Because of the linear relationship between $\mu$ and $V$ in 2D, given by $\mu(V)=V\pi\hbar^2\kappa/(m_+^*+m_-^*)$ \cite{supplmat}, and since $\sigma(V)=n(V) e^2 \tau/m^* $, we find that the normalized $\gamma^\pm(\mu)$ coefficient in Eq. (\ref{eq.polar.emcha.coeff}) indeed decreases following the scaling $\gamma^\pm(V)\sim G(V)/\sigma(V)\sim V^{-5/2}$, as shown in Fig. \ref{fig:rectification.angle}. This behavior results from the Fermi surface (FS) moving away from the Weyl node, and towards regions in the Hilbert space of increasingly Euclidean character, as shown in Fig. \ref{fig:experimental.setup}(b).  

{\it Conclusions --} 
We have shown that the electric magnetochiral anisotropy owes its existence to the quantum geometric properties of noncentrosymmetric media featured by large Born effective charges. We have laid Rikken´s conjecture on solid theoretical ground and we unveiled a universal scaling relationship between the eMChA coefficient and the chemical potential, $\gamma^{\pm}(\mu)\sim \mu^{-5/2}$ \cite{supplmat}, which we verified experimentally through phase sensitive transport measurements in 2D tellurium flakes under applied gate voltage, $V$ \cite{Chang_2023}. Our work further demonstrates the full control of rectification by varying ${\bf I}$, ${\bf P}$, ${\bf B}$, and $V$, opening up new venues for the understanding of nonreciprocal phenomena in advanced quantum materials and paving the way for the design, production, and control of electronic devices based on rectification \cite{Dalvin}.

\begin{acknowledgments}
P.F. acknowledges support of the program Investigo (ref. 200076ID6/BDNS 664047), funded by the European Union through the Recovery, Transformation and Resilience Plan NextGenerationEU. V.V. acknowledges financial support of PNRR MUR project PE0000023-NQSTI and PRIN 2022 (Prot. 20228YCYY7). This work is supported by the Brazilian funding agencies FAPERJ and CNPq. 
\end{acknowledgments}

\bibliography{biblio}

\end{document}


\title{Supplemental Material to 
\texorpdfstring{\\}{}
``Quantum geometry and the electrical magnetochiral anisotropy in noncentrosymmetric polar media"}

\author{Pierpaolo Fontana}
\affiliation{Departament de Física, Universitat Autònoma de Barcelona, 08193 Bellaterra, Spain}

\author{Victor Velasco}
\affiliation{International School for Advanced Studies (SISSA), Via Bonomea 265, I-34136 Trieste, Italy}

\author{Chang Niu}
\affiliation{Elmore Family School of Electrical and Computer Engineering, Purdue University, West Lafayette, Indiana 47907, United States} 
\affiliation{Birck Nanotechnology Center, Purdue University, West Lafayette, Indiana 47907, United States}

\author{Peide D. Ye}
\affiliation{Elmore Family School of Electrical and Computer Engineering, Purdue University, West Lafayette, Indiana 47907, United States} 
\affiliation{Birck Nanotechnology Center, Purdue University, West Lafayette, Indiana 47907, United States}

\author{Pedro V. Lopes}
\affiliation{Instituto de F\'isica, Universidade Federal do Rio de Janeiro, Caixa Postal 68528, 21941-972 Rio de Janeiro, Brazil}

\author{Kaio E. M. de Souza}
\affiliation{Universidade Federal do Rio de Janeiro - Campus UFRJ Duque de Caxias, Rod. Washington Luiz, 19593 - km 104,5 - Santa Cruz da Serra, Duque de Caxias, 25240-005, Rio de Janeiro, Brazil}

\author{Marcus V. O. Moutinho}
\affiliation{Universidade Federal do Rio de Janeiro - Campus UFRJ Duque de Caxias, Rod. Washington Luiz, 19593 - km 104,5 - Santa Cruz da Serra, Duque de Caxias, 25240-005, Rio de Janeiro, Brazil}

\author{Caio Lewenkopf}
\affiliation{Instituto de F\'isica, Universidade Federal do Rio de Janeiro, Caixa Postal 68528, 21941-972 Rio de Janeiro, Brazil}

\author{Marcello B. Silva Neto}
\affiliation{Instituto de F\'isica, Universidade Federal do Rio de Janeiro, Caixa Postal 68528, 21941-972 Rio de Janeiro, Brazil}

\date{\today}

\maketitle


\section{The conduction band $-$ ${\bf k}\cdot{\bf p}$ perturbation theory}

The conduction band structure of tellurium (Te) has been derived using ${\bf k}\cdot{\bf p}$ perturbation theory \cite{Magnetooptical}. We assume that the  spectrum, 
$E_n({\bf k}_H)$, and wave functions, $\psi_{n,{\bf k}_H}({\bf r})$, for the $n-$th conduction band at ${\bf k}_H=(4\pi/3a,0,\pi/c)$, are solutions 
to the eigenvalue problem ${\cal H}_0\psi_{n,{\bf k}_H}({\bf r})=E_n({\bf k}_H)\psi_{n,{\bf k}_H}({\bf r})$, for an unperturbed Hamiltonian
\begin{equation}
{\cal H}_0=\frac{\mathbf{p}^2}{2m}+U(\mathbf{r}),
\end{equation}
where $U(\mathbf{r})$ is the periodic potential. To determine the solutions for ${\bf k}$ away from the $H-$point to the Schrödinger's equation, ${\cal H}\psi_{n,{\bf k}}({\bf r})=E_n({\bf k})\psi_{n,{\bf k}}({\bf r})$, a double perturbation theory in ${\bf k}\cdot{\bf p}$ and the spin-orbit-interaction (SOI) is applied. Substituting $\psi_{n,{\bf k}}({\bf r})=e^{i{\bf k}\cdot{\bf r}}u_{n,{\bf k}_H,{\bf k}}({\bf r})$ into Schrödinger's equation yields an eigenvalue problem for $u_{n,{\bf k}_H,{\bf k}}({\bf r})$:
\begin{equation}
    {\cal H}_{kp,so}u_{n,{\bf k}_H,{\bf k}}({\bf r})=E^\prime_n({\bf k}_H+{\bf k})u_{n,{\bf k}_H,{\bf k}}({\bf r}),
\end{equation}
where $E^\prime_n({\bf k}_H+{\bf k})=E_n({\bf k}_H+{\bf k})-\hbar^2{\bf k}^2/2m$. The matrix elements of ${\cal H}={\cal H}_0+{\cal H}_1+{\cal H}_2+{\cal H}_3$, with 
\begin{eqnarray}
{\cal H}_1&=&\frac{\hbar}{m}{\bf k}\cdot{\bf p},\nonumber\\
{\cal H}_2&=&\frac{\hbar}{4m^2c^2}(\mathbf{\sigma}\times\nabla U(\mathbf{r}))\cdot\mathbf{p},\nonumber\\ 
{\cal H}_3&=&\frac{\hbar^2}{4m^2c^2}(\mathbf{\sigma}\times\nabla U(\mathbf{r}))\cdot\mathbf{k},
\end{eqnarray}
are computed in terms of the unperturbed states $\psi_{n,{\bf k}_H}({\bf r})$, around ${\bf k}={\bf k}_H$. These terms correspond, respectively, to contributions from ${\bf k}\cdot{\bf p}$, from the ${\bf k}-$independent SOI, and from the ${\bf k}-$SOI.

The above procedure leads to a $2\times 2$ Hamiltonian 
\begin{equation}
{\cal H}_c=
\begin{bmatrix}
A k_z^2 + B k_\perp^2 + S k_z & C k_- + iG k_- k_z + F k_+^2 \\
C k_+ - iG k_+ k_z + F k_-^2 & A k_z^2 + B k_\perp^2 - S k_z
\end{bmatrix}.\nonumber
\label{Hamiltonian-Conduction-Band}
\end{equation}
Up to second order in ${\bf k}\cdot{\bf p}$, the Hamiltonian includes the $S k_z$ term, a linear-in-$k_z$ contribution due to the lack of inversion symmetry, and the parabolic $A k_z^2 + B k_\perp^2$ contributions. The combination of ${\bf k}\cdot{\bf p}$ and the spin-orbit interaction gives rise to the off-diagonal, linear-in-$k_\pm$ band mixing elements $C k_\pm$, generating a principal Weyl node at the $H-$point. We also include a trigonal warping term $F k_\pm^2$, responsible for the generation of satellite Weyl nodes away from the $H-$point, and a term arising from the ${\bf k}-$dependent spin-orbit interaction, i.e., $\pm iG k_\mp k_z$. 


\section{The perturbed distribution function}

Let us calculate the current density
\begin{equation}
    {\bf j}=-e\sum_{n,{\bf k}}\dot{\bf r}_n f_n({\bf k})=
    -e\sum_{n,{\bf k}}\dot{\bf r}_n\,\delta f_n({\bf k}),
\end{equation}
with ${\bf j}=-e\sum_{n,{\bf k}}\dot{\bf r}_n f_n^0({\bf k})=0$, but in the presence of both external electric field, ${\bf E}$, and magnetic field, ${\bf B}$. Here $f_n({\bf k})=f_n^0({\bf k})+\delta f_n({\bf k})$ is the {\it non-equilibrium distribution function} satisfying the Boltzmann transport equation in the relaxation time approximation
\begin{equation}
        \frac{\partial f_n({\bf k})}{\partial t}+\dot{\bf r}_n\cdot\frac{\partial f_n({\bf k})}{\partial {\bf r}}+\dot{\bf k}_n\cdot\frac{\partial f_n({\bf k})}{\partial {\bf k}}=-\frac{f_n({\bf k})-f_n^0({\bf k})}{\tau_{\bf k}},
\end{equation}   
where $f_n^0({\bf k})=1/(e^{\beta(\varepsilon_n({\bf k})-\mu)}+1)$ is the equilibrium distribution, $\varepsilon({\bf k})$ is the dispersion, $\beta=1/k_B T$ is the inverse temperature, $\mu$ is the chemical potential and $\tau_{\bf k}=\tau$ is the relaxation time. We shall consider a homogeneous system, $\partial f/\partial {\bf r}=0$ and the DC limit, where $\partial f/\partial t=0$. In this case, the Boltzmann transport equation simplifies to
\begin{equation}
        \dot{\bf k}_n\cdot\frac{\partial f_n({\bf k})}{\partial {\bf k}}=-\frac{\delta f_n({\bf k})}{\tau},
\end{equation}   
and our goal is to calculate the perturbation of the distribution function, $\delta f_n({\bf k})$.

The acceleration $\dot{\bf k}_n$ is defined by the semiclassical equations of motion (we shall henceforth use $c=1$)
\begin{equation}
    \hbar\dot{\bf k}_n=-e({\bf E}+\dot{\bf r}_n\times{\bf B}),
    \label{Acceleration}
\end{equation}
and the origin of each term is described in the manuscript.

\subsection{Electric field}

Let us first consider the case of an applied electric field, ${\bf E}\neq 0$, in the absence of a magnetic field, i.e., ${\bf B}=0$. In this case, the Boltzmann equation is given by
\begin{equation}
        \left(\frac{-e{\bf E}}{\hbar}\right)\cdot\frac{\partial f_n}{\partial {\bf k}}=-\frac{\delta f_n}{\tau}.
\end{equation}   
If we use $f_n=f_n^0$ on the left-hand side and recall that
\begin{equation}
    \frac{\partial f_n^0}{\partial {\bf k}}=
    \frac{\partial f_n^0}{\partial \varepsilon_n}
    \frac{\partial \varepsilon_n}{\partial {\bf k}}=
    \hbar{\bf v}^n_{\bf k}\frac{\partial f_n^0}{\partial \varepsilon},
\end{equation}
then we can write the perturbation as
\begin{equation}
    \delta f_n=e\tau\,{\bf v}^n_{\bf k}\cdot{\bf E}
    \left(\frac{\partial f_n^0}{\partial \varepsilon_n}\right).
\end{equation}

\subsection{Including the magnetic field}

Let us now consider the case of applied electric, ${\bf E}\neq 0$, and magnetic, ${\bf B}\neq 0$, fields. In this case, the Boltzmann equation is given by
\begin{equation}
        \left(\frac{e{\bf E}}{\hbar}\right)\cdot\frac{\partial f_n}{\partial {\bf k}}+\left(\frac{e(\dot{\bf r}_n\times{\bf B})}{\hbar}\right)\cdot\frac{\partial f_n}{\partial {\bf k}}=\frac{\delta f_n}{\tau}.
\end{equation}   
If we now try to replace $f_n$ in $\partial f_n/\partial {\bf k}_n$ by the equilibrium distribution $f_n^0$ and also replace $\dot{\bf r}_n$ by ${\bf v}^n_{\bf k}$, then the result is that the magnetic field drops out. In fact, we get
\begin{equation}
    \left(\frac{e({\bf v}^n_{\bf k}\times{\bf B})}{\hbar}\right)\cdot\frac{\partial f_n^0}{\partial {\bf k}}=\left(\frac{e({\bf v}^n_{\bf k}\times{\bf B})}{\hbar}\right)\cdot{\bf v}^n_{\bf k}\frac{\partial f_n}{\partial \varepsilon_n}=0,
\end{equation}
because of the vector product $({\bf v}^n_{\bf k}\times{\bf B})\cdot{\bf v}^n_{\bf k}=0$. We need to keep the non-equilibrium distribution in the part related to the magnetic field. Let us call for simplicity 
\begin{equation}
    g^n=f_n-f_n^0,
\end{equation}
and treat ${\bf B}$ as a perturbation, in which case
\begin{equation}
    g^n=g^n_0+g^n_1+g^n_2+\cdots,
\end{equation}
where $g_m\sim O(B^m)$. If we are interested in contributions up to order ${\bf B}$, then the Boltzmann equation reduces to
\begin{equation}
        -e\left({\bf v}^n_{\bf k}\cdot{\bf E}\right)\cdot\frac{\partial f_n^0}{\partial \varepsilon_n}-e\frac{(\dot{\bf r}_n\times{\bf B})}{\hbar}\cdot\frac{\partial (g^n_0+g^n_1))}{\partial {\bf k}}=-\frac{g^n_0+g^n_1}{\tau}.
\end{equation}   
The left-hand side and the right-hand side must be equal, order by order, in the powers of the magnetic field ${\bf B}$. To zeroth order, we have
\begin{equation}
    g^n_0=e\tau\,{\bf v}^n_{\bf k}\cdot{\bf E}
    \left(\frac{\partial f_n^0}{\partial \varepsilon_n}\right),
    \label{eq.g0}
\end{equation}
which is the same result we have already obtained. To first order in ${\bf B}$ we are left with
\begin{equation}
        -e\frac{(\dot{\bf r}_n\times{\bf B})}{\hbar}\cdot\frac{\partial g^n_0}{\partial {\bf k}}=-\frac{g^n_1}{\tau},
\end{equation}
since terms such as ${\bf B} g^n_1$ are of order ${\bf B}^2$. In this case, we can calculate
\begin{equation}
    g^n_1 = e\tau\,\frac{(\dot{\bf r}_n\times{\bf B})}{\hbar}\cdot\frac{\partial g^n_0}{\partial {\bf k}}=
    e^2\tau^2\,\frac{(\dot{\bf r}_n\times{\bf B})}{\hbar}\cdot
    \nabla_{\bf k}(\dot{\bf r}^n\cdot{\bf E})
    \left(\frac{\partial f_n^0}{\partial \varepsilon_n}\right),
    \label{eq.g1}
\end{equation}
which is precisely Eq. (12) in the manuscript.

\section{Geodesic contribution to the velocity}

Let us now include in the expression for $\delta f_n=g^n_0+g^n_1$ given in eqs. (\ref{eq.g0}) and (\ref{eq.g1}) the geodesic contribution to the velocity of the wave packet \cite{Gravity}, as well as the existence of an intrinsic, static, and uniform electric field, $\boldsymbol{\cal E}_0$
\begin{eqnarray}
    \dot{\bf r}_n&=&{\bf v}^n_{\bf k}+
    \dot{\bf k}_n\times{\bf \Omega}^n_{\bf k}+\dot{\bf k}_n\cdot{\bf \Gamma}^n(\partial{\bf g})\cdot\dot{\bf k}_n,\label{Velocity}\\
    \hbar\dot{\bf k}_n&=&-e({\bf E}+\boldsymbol{\cal E}_0)-e\dot{\bf r}_n\times{\bf B}.
\end{eqnarray}
In the equation for $\dot{\bf r}_n$, ${\bf v}^n_{\bf k}=(1/\hbar)\nabla_{\bf k}\varepsilon_n({\bf k})$ represents the $n-$band velocity and $\dot{\bf k}_n\times{\bf \Omega}^n_{\bf k}$ represents the anomalous velocity, arising from the Berry curvature. The last term, with $\dot{\bf k}_n\cdot{\bf \Gamma}^n(\partial{\bf g})\cdot\dot{\bf k}_n\equiv{\bf \Gamma}^n_{ij\ell}\dot{\bf k}^n_j\dot{\bf k}^n_\ell$, is the geodesic contribution from the energy-normalized quantum metric dipole \cite{Kaplan_2024_PRL}, represented by the Christoffel-like symbols ${\bf \Gamma}_{ij\ell}$, defined as
\begin{equation}
\Gamma^n_{ij\ell}=2\partial_{i}\tilde{g}^n_{j\ell}-\frac{1}{2}(\partial_{j}\tilde{g}^n_{\ell i}+\partial_{\ell}\tilde{g}^n_{ji}),
\end{equation}
where 
\begin{equation}
    \tilde{g}^\pm_{ij}=\frac{g^\pm_{ij}}{\varepsilon_\pm-\varepsilon_\mp}.
\end{equation}
The geodesic contribution to the wavepacket velocity of order ${\bf E}$ is given by
\begin{equation}
    \dot{\bf r}_n=2\frac{e^2}{\hbar^2}{\bf E}\cdot{\bf \Gamma}^n(\partial{\bf g})\cdot\boldsymbol{\cal E}_0+\cdots,
\end{equation}
and we have used the fact that $\Gamma_{ijk}$ is symmetric in the $j,k$ indices such that $\Gamma_{ijk}E_j{\cal E}_{0,k}=\Gamma_{ijk}{\cal E}_{0,j}E_k$, which provides the factor $2$ in the above equation. Now, the contribution of order $\bf{E\cdot B}$ for $\delta f$ from the band $n$ becomes
\begin{equation}
    \delta f = 2 \frac{e^4\tau^2}{\hbar^3}
    [({\bf E}\cdot{\bf \Gamma}^n(\partial{\bf g})\cdot\boldsymbol{\cal E}_0)\times{\bf B}]\cdot\nabla_{\bf k}({\bf v}^n_{\bf k}\cdot{\bf E})
    \left(\frac{\partial f_0}{\partial \varepsilon_n}\right)+\cdots.
    \label{eq.deltaf}
\end{equation}
In terms of the $\delta f$ calculated above, the geodesic contribution to the total current density is
\begin{widetext}
\begin{equation}
\delta{\bf j}=\frac{2e^5\tau^2}{\hbar^3}\sum_{n,{\bf k}}{\bf v}^n_{\bf k}
\left\{[({\bf E}\cdot{\bf \Gamma}^n(\partial{\bf g})\cdot\boldsymbol{\cal E}_0)\times{\bf B}]\cdot\nabla_{\bf k}({\bf v}^n_{\bf k}\cdot{\bf E})\right\}
    \left(-\frac{\partial f_0}{\partial \varepsilon_n}\right),
\end{equation}
The above quantity can be expressed in components as
\begin{eqnarray}
\delta {\bf j}_i&=&C\sum_{n,{\bf k}}{v}^n_{{\bf k},i}
    \left\{[({\bf E}\cdot{\bf \Gamma}^n(\partial{\bf g})\cdot\boldsymbol{\cal E}_0)\times{\bf B}]_a (\partial_a {v}^n_{{\bf k},j}) {E}_j\right\}
    \left(-\frac{\partial f_0}{\partial \varepsilon_n}\right),\nonumber\\
    &=&C\sum_{n,{\bf k}}{v}^n_{{\bf k},i}
    \left\{\epsilon_{ab\ell}({\bf E}\cdot{\bf \Gamma}^n(\partial{\bf g})\cdot\boldsymbol{\cal E}_0)_b{B}_\ell (\partial_a {v}^n_{{\bf k},j}) {E}_j\right\}
    \left(-\frac{\partial f_0}{\partial \varepsilon_n}\right)\nonumber\\
    &=&C\sum_{n,{\bf k}}{v}^n_{{\bf k},i}
    \left\{\epsilon_{ab\ell}{\Gamma}^n_{bkm}(\partial{\bf g}){\bf E}_k\boldsymbol{\cal E}_{0,m}{B}_\ell (\partial_a {v}^n_{{\bf k},j}){E}_j\right\}
    \left(-\frac{\partial f_0}{\partial \varepsilon_n}\right)\nonumber,
\end{eqnarray}
\end{widetext}
where we defined $C=2e^5\tau^2/\hbar^3$ to shorten the notation. The above expression can be compactly written as 
\begin{equation}
\delta {\bf j}_i=G_{ijk\ell}E_j E_k B_\ell
\end{equation}
in terms of a $4-$rank tensor $G_{ijk\ell}$
\begin{equation}
    G_{ijk\ell}=\frac{2e^5\tau^2}{\hbar^3}
    \sum_{n,{\bf k}}{v}^n_{{\bf k},i}
    \left\{\epsilon_{ab\ell}{\Gamma}^n_{bkm}(\partial{\bf g})\boldsymbol{\cal E}_{0,m}\right\}(\partial_a {v}^n_{{\bf k},j})
    \left(-\frac{\partial f_0}{\partial \varepsilon_n}\right).
    \label{eq.G-tensor}
\end{equation}
The geometry of our problem is such that
\begin{eqnarray}
    \boldsymbol{\cal E}_0&\parallel&{\bf x},\nonumber\\
    {\bf B}&\parallel&{\bf y},\nonumber\\
    {\bf E}&\parallel&{\bf z},\nonumber\\
    {\bf j}, {\bf v}&\parallel&{\bf z}.
    \label{eq.geometry}
\end{eqnarray}
According to the geometry in Eq. (\ref{eq.geometry}) and the expression for $\delta {\bf j}$ we need to evaluate the $G_{zzzy}$ component of the tensor in Eq. (\ref{eq.G-tensor}). If we recall that $\epsilon_{ab\ell}$ is the completely antisymmetric tensor, then for $a=z$ and $\ell=y$ the only possibility left for $b=x$, such that $\epsilon_{zxy}\neq 0$. Moreover, since $\boldsymbol{\cal E}_0\parallel{\bf x}$, $m=x$ and the expression of $G_{zzzy}$ becomes
\begin{equation}
    G_{zzzy}(\mu)=\frac{2e^5\tau^2}{\hbar^3}
    \sum_{n,{\bf k}}{v}^n_{{\bf k},z}
    \left\{{\Gamma}^n_{xzx}(\partial{\bf g})\boldsymbol{\cal E}_{0,x}\right\}
    (\partial_z {v}^n_{{\bf k},z})\delta(\varepsilon_n({\bf k})-\mu).
    \label{eq.G-tensor-zzzy}
\end{equation}
In the above expression, the Levi-Civita symbol is such that $\epsilon_{xyz}=+1$, and we have also used that at low temperatures $(-\partial f_0/\partial\varepsilon)=\delta(\varepsilon({\bf k})-\mu)$, where $\mu$ is the chemical potential.

\subsection{Evaluation of $G_{zzzy}$}

\begin{figure}[t]
    \includegraphics[scale=0.35]{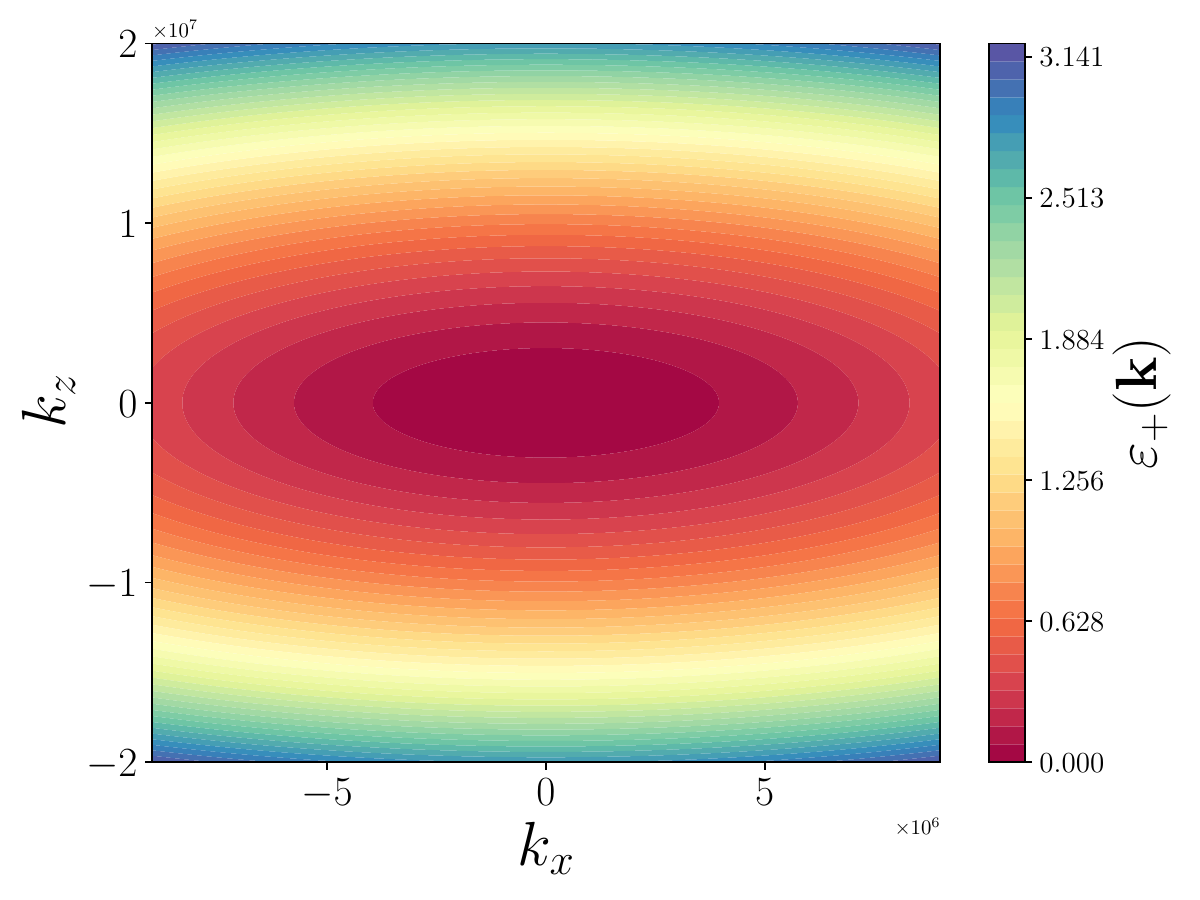}
    \qquad
    \includegraphics[scale=0.35]{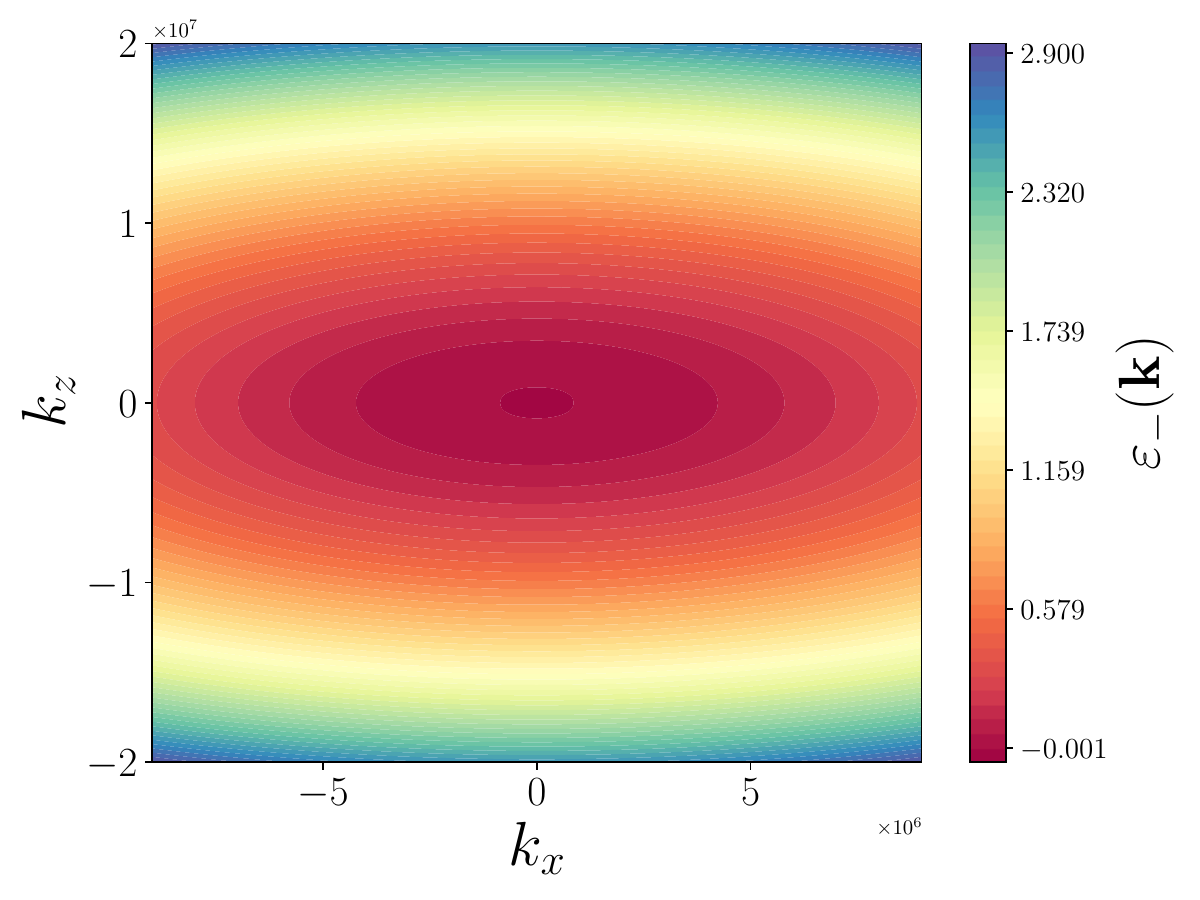}
    \caption{Contour plots, at fixed $k_y=0$, for the conduction bands $\varepsilon_\pm({\bf k})$ for a single Weyl node, in the $k_x-k_z$ plane.}
    \label{fig:conduction-bands}
\end{figure}

For simplicity, let us neglect both the trigonal warping, $F=0$, as well as the ${\bf k}-$dependent spin-orbit interaction, $G=0$, which are the smallest parameters in the problem. In this case, the conduction bands have dispersion relations
\begin{equation}
    \varepsilon_\pm({\bf k})=A k_z^2 + B k_\perp^2 \pm \sqrt{S^2 k_z^2+C^2 k_\perp^2}.
    \label{Two-Bands}
\end{equation}
The expression for the $z-$component of the velocity, $v^\pm_z({\bf k})$, entering in Eq. (\ref{eq.G-tensor-zzzy}), for the two bands, and the $xzx$ component of the Christoffel symbol, $\Gamma_{xzx}({\bf k})$, can also be found analytically
\begin{eqnarray}
    v^\pm_z({\bf k})&=&2A k_z\pm\frac{k_z S^2}{\sqrt{S^2 k_z^2+C^2 k_\perp^2}}\nonumber\\
    \Gamma^\pm_{xzx}({\bf k})&=&\pm\frac{5 S^2 C^4 k_x^2 k_z}{8(C^2 k_\perp^2+S^2k_z^2)^{7/2}}.
\end{eqnarray}
Then the expression for $G_{zzzy}$ in Eq. (\ref{eq.G-tensor-zzzy}) can be computed analytically in the parabolic approximation, where $\varepsilon_\pm({\bf k})=\hbar^2 {\bf k}^2/2m_\pm^*$ for both bands with $m^*_+\neq m^*_-$ at the same chemical potential $\mu$, as shown in Fig. \ref{fig:conduction-bands}.

\begin{figure}[t]
\includegraphics[scale=0.6]{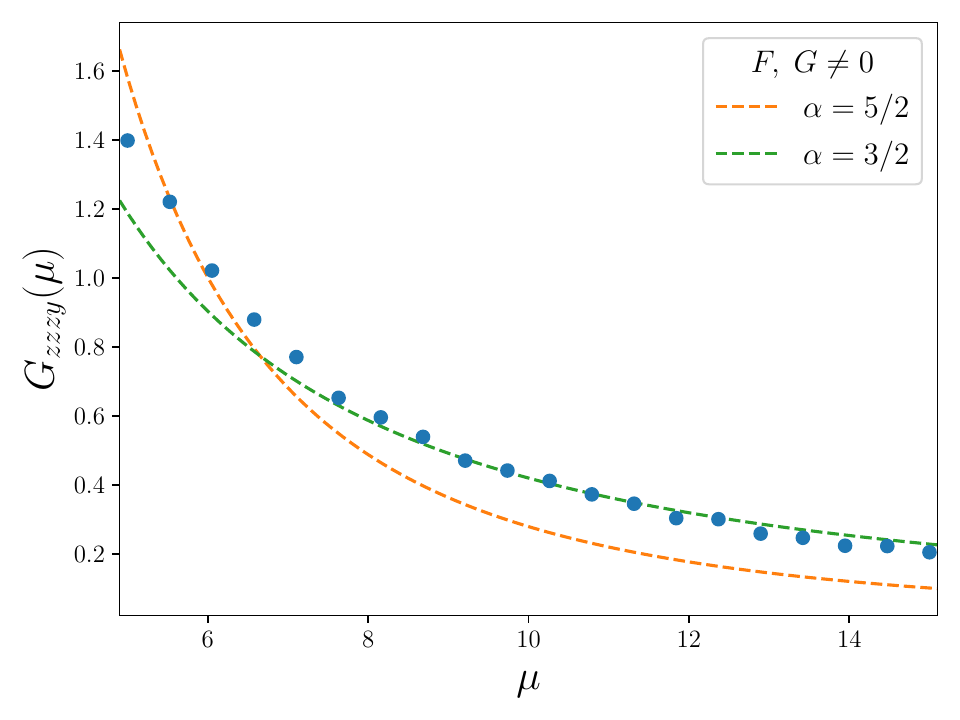}
\caption{Theory vs numerics -- the $\mu^{-3/2}$ scaling. The blue markers represent the numerical results for the full dependence of the $G_{ijk\ell}(\mu)$ tensor on the chemical potential, $\mu$, including the contributions from trigonal warping, $F\neq 0$, and $k-$dependent spin-orbit interaction, $G\neq 0$. The two dashed lines represent two different scaling regimes, $G(\mu)\sim \mu^{-\alpha}$. The green curve is calculated in the parabolic approximation, and with $F=G=0$, for which $\alpha=3/2$. The parabolic approximation works very well away from the Weyl nodes, and deviates only slightly at small values of the chemical potential, for which we expect $\alpha=5/2$ (orange curve).}
\label{fig:theo-numerics}
\end{figure}

For a two-dimensional system confined to the $z-x$ plane, where $k_y=0$, we can use polar coordinates 
\begin{eqnarray}
    k_x &=& k\sin(\alpha),\nonumber\\
    k_z &=& k\cos(\alpha),  
\end{eqnarray}
where $\alpha$ is the angle between the ${\bf k}-$vector and the $k_z-$axis. Replacing now $\sum_{\bf k}\rightarrow {\cal V}\int d^2{\bf k}/(2\pi)^2$, where ${\cal V}$ is the volume of the 2D Brillouin zone, and collecting the  parabolic contribution to $G_{zzzy}$ we obtain
\begin{widetext}
\begin{eqnarray}
    G^\pm_{zzzy}(\mu)&=&\frac{2 e^5\tau^2\boldsymbol{\cal E}_{0,x}}{\hbar^3}
    {\cal V}\int\frac{d^2{\bf k}}{(2\pi)^2}
    \underbrace{\left(\frac{\hbar k_z }{m^*_\pm}\right)}_{\text{from ${v}^n_{{\bf k},z}$}}
    \underbrace{\left[\frac{(\pm)5 S^2 C^4k_x^2k_z}{8(C^2 k_x^2+S^2k_z^2)^{7/2}}\right]}_{\text{from $\Gamma_{xzx}({\bf k})$}}
    \underbrace{\left(\frac{\hbar}{m^*_\pm}\right)}_{\text{from $\partial_z{v}^n_{{\bf k},z}$}}
    \delta(\varepsilon_\pm({\bf k})-\mu),\nonumber\\
    &=&\pm\frac{2 e^5\tau^2\boldsymbol{\cal E}_{0,x}}{\hbar^3}
    {\cal V}\frac{\hbar^2}{(m^*_\pm)^2} \int\frac{d^2{\bf k}}{(2\pi)^2}
    \frac{5 S^2 C^4k_x^2k_z^2}{8(C^2 k_x^2+S^2k_z^2)^{7/2}}
    \delta(\varepsilon_\pm({\bf k})-\mu),\nonumber\\
    &=&\pm\frac{2 {\cal V} e^5\tau^2\boldsymbol{\cal E}_{0,x}}{(2\pi)^2\hbar^3}
    \frac{\hbar^2}{(m^*_\pm)^2}\int_0^{2\pi}d\alpha\int\, dk\, k
    \frac{5 S^2 C^4k^2\sin^2(\alpha)\; k^2\cos^2(\alpha)}{8(C^2 k^2\sin^2(\alpha)+S^2 k^2\cos^2(\alpha))^{7/2}}\delta\left(\frac{\hbar^2 k^2}{2m_\pm^*}-\mu\right),\nonumber\\
    &=&\pm\frac{4 m_\pm^* {\cal V} e^5\tau^2\boldsymbol{\cal E}_{0,x}}{(2\pi)^2\hbar^5}\frac{\hbar^2}{(m^*_\pm)^2}
    \int_0^{2\pi}d\alpha\int\, \frac{dk}{k^2}\, 
    \frac{5 S^2 C^4\sin^2(\alpha)\; \cos^2(\alpha)}{8(C^2 \sin^2(\alpha)+S^2 \cos^2(\alpha))^{7/2}}\delta\left(k^2-\frac{2 m_\pm^*\mu}{\hbar^2}\right),\nonumber\\
    &=&\pm\frac{4 m_\pm^* {\cal V} e^5\tau^2\boldsymbol{\cal E}_{0,x}\,{\cal I}(C,S)}{(2\pi)^2\hbar^5}\frac{\hbar^2}{(m^*_\pm)^2}
    \int\, \frac{dk}{k^2}\, \frac{\hbar}{2\sqrt{2m_\pm^*\mu}}
    \delta\left(k-\frac{\sqrt{2 m_\pm^*\mu}}{\hbar}\right),\nonumber\\
    &=&\pm\frac{2 m_\pm^* {\cal V} e^5\tau^2\boldsymbol{\cal E}_{0,x}\,{\cal I}(C,S)}{(2\pi)^2\hbar^5}\frac{\hbar^2}{(m^*_\pm)^2}
    \frac{\hbar^2}{2m_\pm^*\mu}\frac{\hbar}{\sqrt{2m_\pm^*\mu}},\nonumber\\
    &=&\pm\frac{{\cal V} e^5\tau^2\boldsymbol{\cal E}_{0,x}\,{\cal I}(C,S)}{(2\pi)^2 \sqrt{2} (m_\pm^*)^{5/2}}\frac{1}{\mu^{3/2}},
\end{eqnarray}
\end{widetext}
where we have used that 
\begin{equation}
\delta({\bf k}^2-k_F^2)=\frac{1}{2 k_F}\left\{\delta(k-k_F)+\delta(k+k_F)\right\},
\end{equation}
with $k_F=\sqrt{2m^*\mu}/\hbar$ and we have also defined the angular integral
\begin{equation}
    {\cal I}(C,S)=\int_0^{2\pi}d\alpha 
    \frac{5S^2 C^4\sin^2(\alpha)\; \cos^2(\alpha)}{8(C^2 \sin^2(\alpha)+S^2 \cos^2(\alpha))^{7/2}}.
\end{equation}
We conclude that within the parabolic approximation, $G_{zzzy}$ should scale with the chemical potential $\mu$ according to the relation
\begin{equation}
    G_{zzzy}(\mu)=\frac{G_0}{\mu^{3/2}},
\end{equation}
where 
\begin{equation}
    G_0=\frac{{\cal V} e^5\tau^2\boldsymbol{\cal E}_{0,x}\,{\cal I}(C,S)}{(2\pi)^2\sqrt{2}} \left(\frac{1}{(m_+^*)^{5/2}}-\frac{1}{(m_-^*)^{5/2}}\right).
\end{equation}
In Fig. \ref{fig:theo-numerics} we show the comparison between our exact calculation for the tensor in \eqref{eq.G-tensor-zzzy} and the parabolic approximation.

\section{Universal scaling $-$ $G(\mu)\sim\mu^{(d-5)/2}$}

Let us now demonstrate that the scaling $G(\mu)\sim\mu^{-3/2}$ obtained above is simply an example of a more general and universal scaling, valid in any dimensionality $d$. For any parabolic band that is disturbed by a Weyl node
\begin{equation}
    {\cal H}={\bf k}^2\,\sigma_0+{\bf k}\cdot\sigma,
    \label{eq.toy-Hamiltonian}
\end{equation}
the eigenvalues are
\begin{equation}
    \varepsilon_\pm({\bf k})={\bf k}^2\pm |{\bf k}|.
\end{equation}
The quantity we want to evaluate is
\begin{equation}
    G(\mu)\propto
    \sum_{{\bf k}}{\bf v}_{{\bf k}}\,
    \partial{\bf g}({\bf k})\,
    \partial {\bf v}_{{\bf k}}\,
    \delta(\varepsilon({\bf k})-\mu),
    \label{eq.G-scaling}
\end{equation}
where ${\bf v}_{{\bf k}}\sim\partial\varepsilon({\bf k})$ and ${\bf g}({\bf k})={\bf G}({\bf k})/\Delta\varepsilon({\bf k})$ is the energy normalized quantum metric with
\begin{equation}
\Delta\varepsilon({\bf k})\equiv\varepsilon_+({\bf k})-\varepsilon_-({\bf k})=
2|{\bf k}|=2({\bf k}^2)^{1/2}.
\end{equation}
For the case of the Hamiltonian (\ref{eq.toy-Hamiltonian}) the velocity and the quantum metric read
\begin{eqnarray}
    {\bf v}_{{\bf k}}&=& 2{\bf k}\pm\frac{{\bf k}}{({\bf k}^2)^{1/2}},\nonumber\\
    {\bf G}_{ij}({\bf k})&=& \frac{{\bf k}^2\delta_{ij}-k_ik_j}{({\bf k}^2)^2},\nonumber\\
    {\bf g}_{ij}({\bf k})&=& \pm\frac{{\bf k}^2\delta_{ij}-k_ik_j}{2({\bf k}^2)^{5/2}}.
\end{eqnarray}
The parabolic approximation implies that
\begin{equation}
    {\bf v}_{{\bf k}}\approx2{\bf k}.
\end{equation}
Now it is a matter of analyzing the scaling of the quantities in eq. (\ref{eq.G-scaling}) with respect to $|{\bf k}|=k$. For both bands $n=\pm$, we have that
\begin{eqnarray}
    \left[{\bf v}^n_{{\bf k}}\right]_k &\sim & k,\nonumber\\
    \left[\partial{\bf g}_n({\bf k})\right]_k &\sim & \frac{k^3}{k^{7}}\sim\frac{1}{k^4},\nonumber\\
    \left[\partial{\bf v}^n_{{\bf k}}\right]_k &\sim & k^0,\nonumber\\   
    \left[\sum_{{\bf k}}\right]_k &\sim & k^{d-1},\nonumber\\
    \left[\delta(\varepsilon_n({\bf k})-\mu)\right]_k &\sim & \frac{1}{k}.
\end{eqnarray}
We conclude that $\left[G\right]_k\sim k^{d-5}$ and since we are far away from the Weyl nodes, where the parabolic approximation holds, we have $k\sim\sqrt{\mu}$, we end up with
\begin{equation}
    G(\mu)\sim\mu^{(d-5)/2},
\end{equation}
which, for $d=2$, gives us the $G(\mu)\sim\mu^{-3/2}$ result obtained in the manuscript and shown in Fig. \ref{fig:theo-numerics}. 

\section{Gating and the chemical potential}

\begin{figure}[t]
\includegraphics[scale=0.7]{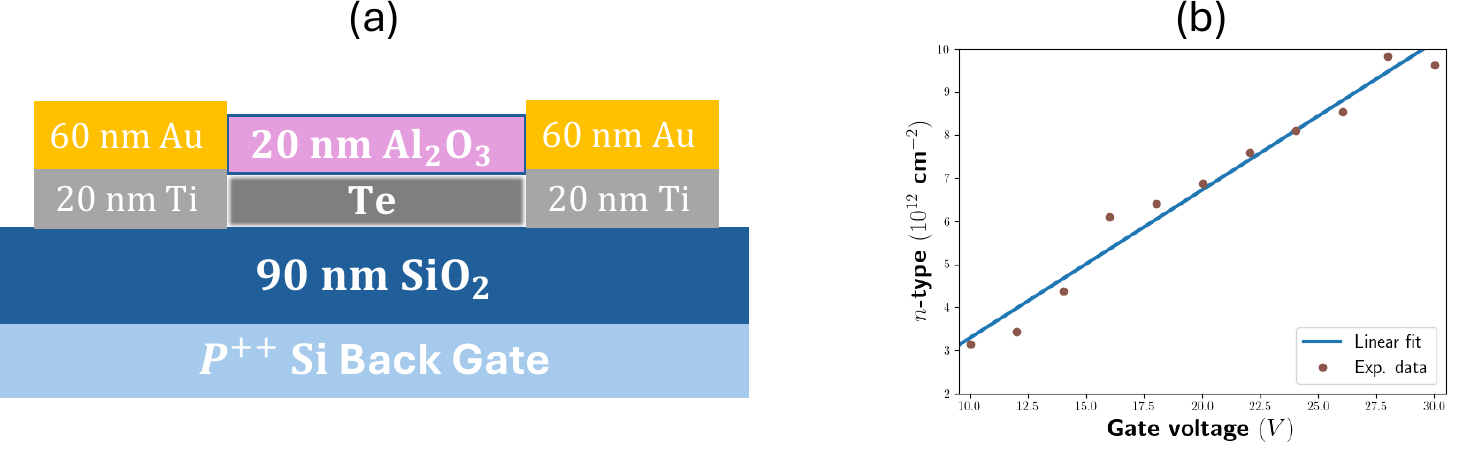}
\caption{(a) Pictorial representation of an $n$-type Te field-effect transistor. (b) Gate-dependent carrier density $n$ (brown circles) extracted from Hall-measurements at $T=1$K.}
\label{fig:Gating-setup}
\end{figure}

A Te device with Hall-bar structure was used for magneto-transport studies. Source and drain electrodes were made along the long edge of the flake, which coincides with the direction of the atomic chains. Hall bars are used to measure the longitudinal and transverse resistance, with the back-gate tuning the film electron density \cite{Chang_2023}. A sketch of a typical $n$-type field-effect device is shown in Fig. \ref{fig:Gating-setup}(a). A $20$nm Al$_2$O$_3$ is grown by atomic layer deposition (ALD) at $200^\circ$C to dope tellurium into $n$-type. Tellurium flakes with a thickness around $10$nm are transferred on to $90$nm SiO$_2$/Si substrate, which serves as a back-gate for the transistor. The layers of $20$nm Ti /$60$nm Au are used for the $n$-type contact. In order to prove the $n$-type doping effect, a transfer curve of a Te field-effect device is measured using a four-terminal method. By carrying out the Hall measurement we can calculate the two-dimensional carrier density $n$ and electron Hall mobility $\mu$ under different gate biases using \cite{Chang_2023}
\begin{eqnarray}
    n&=&\frac{1}{e}\left(\frac{dB}{dR_{xy}}\right),\nonumber\\
    \mu&=&\left(\frac{L}{W}\right)\left(\frac{1}{R_{xx}ne}\right),
\end{eqnarray}
where $L$ is channel length, $W$ is channel width, $R_{xx}$ is longitudinal resistance, and $R_{xy}$ is the Hall resistance. More than 10 devices were fabricated and measured, all of which showed similar and reproducible behaviors. The evolution of the $n-$type carrier density with gating is shown in Fig. \ref{fig:Gating-setup}(b).


For a two-dimensional electron gas (2DEG), the density of states is constant and is solely determined by the effective mass, $m^*$, through the formula
\begin{equation}
    g(E)=\frac{m^*}{\pi\hbar^2}.
\end{equation}
At zero temperature, the Fermi-Dirac distribution simplifies because all states up to the chemical potential (Fermi level, $\mu$) are occupied, and all states above it are unoccupied. The carrier concentration $n$ can be calculated directly from the density of states and the chemical potential
\begin{equation}
    n=\int_0^\mu g(E)dE.
\end{equation}
Since $g(E)$ is constant, the relation between $n$ and $\mu$ reduces to
\begin{equation}
    n(\mu)=\frac{m^*\mu}{\pi\hbar^2}.
\end{equation}
On the other hand, the carrier density exhibits a linear dependence on the gate voltage, as shown in Fig. \ref{fig:Gating-setup}(b), which can be written as
\begin{equation}
    n(V)=\kappa V,\quad\quad\mbox{with}\quad\quad\kappa\approx 0.335\times 10^{12}/(V cm^2).
\end{equation}
We conclude that there is a linear relationship between the chemical potential and the gate voltage described by
\begin{equation}
    \mu(V)=\frac{\pi\hbar^2\kappa}{m^*}V.
\end{equation}

\section{Phenomenology and second harmonic generation}

\begin{figure}[h]
    \includegraphics[scale=0.25]{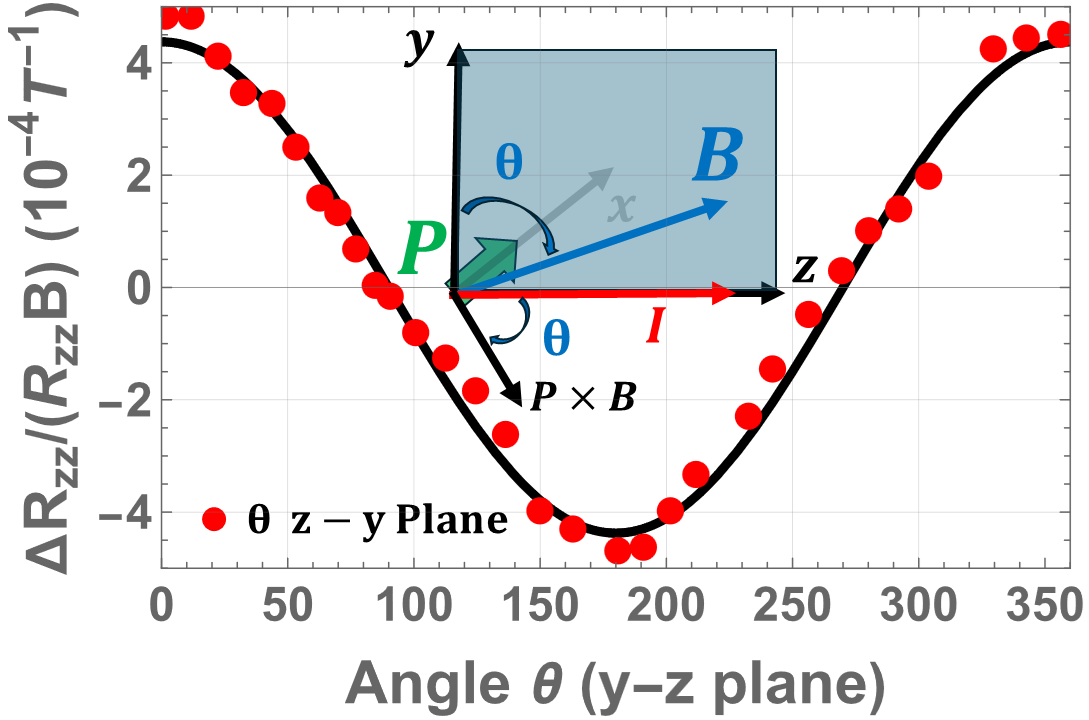}
    \includegraphics[scale=0.25]{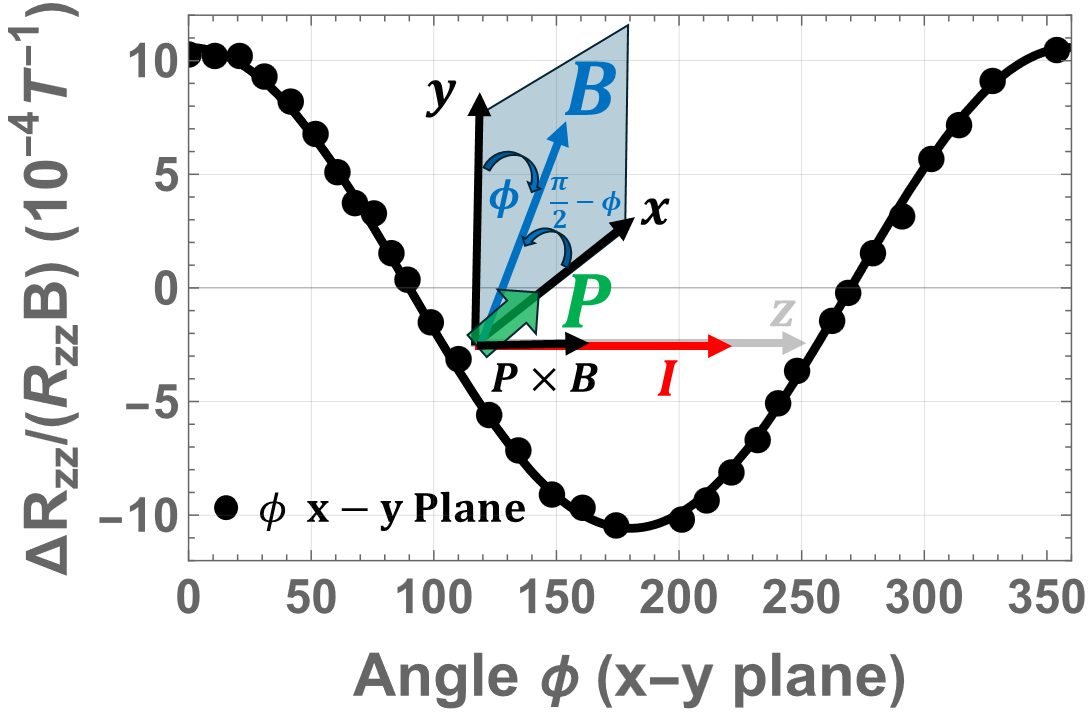}
    \includegraphics[scale=0.26]{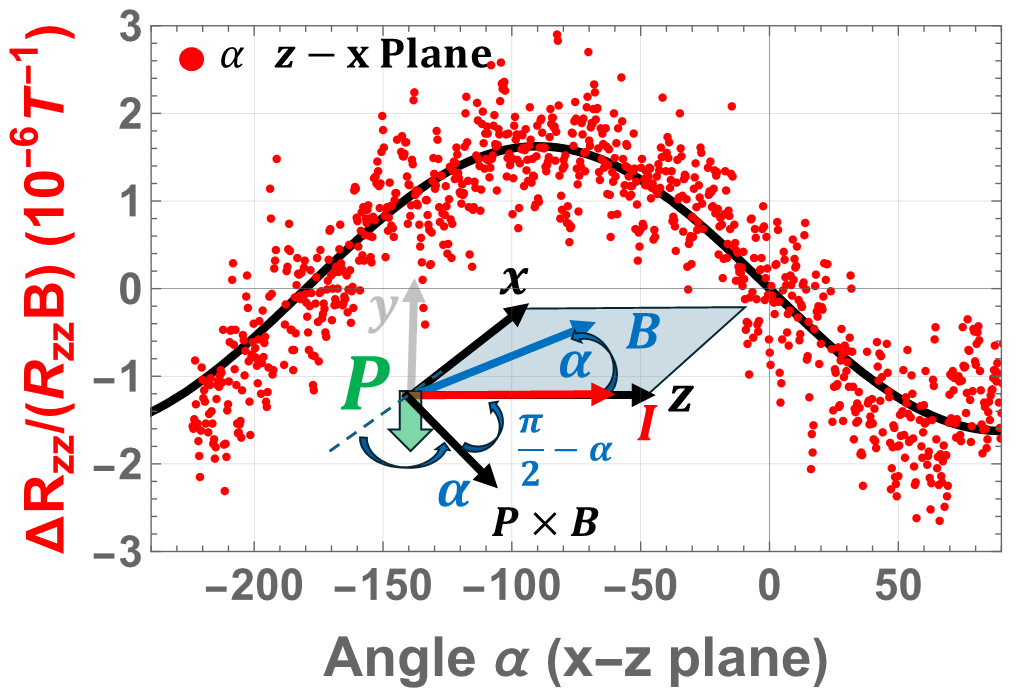}
    \caption{Rotation scans as ${\bf B}$ moves away from: (a) the $\hat{\bf y}-$axis, by an angle $\theta$, and towards the $\hat{\bf z}-$axis, always confined to the $y-z$ plane; (b) away from the $\hat{\bf y}-$axis, by an angle $\phi$, and towards the $\hat{\bf x}-$axis, always confined to the $x-y$ plane; (c) away from the $\hat{\bf z}-$axis, by an angle $\alpha$, and towards the $\hat{\bf x}-$axis, always confined to the $x-z$ plane.}
    \label{fig:rotation-scans}
\end{figure}

In nonreciprocal electrical transport measurement, the AC current with the frequency $\omega$ can be expressed by $I^\omega=I_0\sin(\omega t)$. The unidirectional magnetoresistance is described by the relation
\begin{equation}
R(I,B,P)=R_0[1+\beta B^2 + \gamma^{\pm}{\bf I}\cdot({\bf P}\times{\bf B})].   
\end{equation}
For ${\bf B}\perp{\bf P}\perp{\bf I}$, the nonlinear voltage components can be expressed by 
\begin{equation}
V_{zz}^{2\omega}(t)=\gamma^\pm R_0 B P I_0^2 \sin^2(\omega t)=
\frac{1}{2}\gamma^\pm R_0 B P I_0^2\left[1+\sin\left(2\omega t-\frac{\pi}{2}\right)\right].
\end{equation}
In the experiment, the component of the second harmonic voltage drop detected by the lock-in amplifier is expressed as $V_{zz}^{2\omega}=(1/2)\gamma^\pm R_0 B P I_0^2$, due to its cosine dependence \cite{Chang_2023}. The ordinary magnetoresistance follows the well-known relation $V_{zz}^{\omega}(t)=R_0(1+\beta B^2)I_0\sin(\omega t)$, and the component of the first harmonic voltage drop detected by the lock-in amplifier is expressed as $V_{zz}^{\omega}=R_0(1+\beta B^2)I_0$ \cite{Chang_2023}. If $\beta$ is small, according to the definition, the resistance difference between positive and negative current can be written as
\begin{equation}
    \frac{4 V_{zz}^{2\omega}}{V_{zz}^\omega}=\frac{\Delta R}{R_0}=
    2\gamma^{\pm}{\bf I}\cdot({\bf P}\times{\bf B}), 
\end{equation}
and its evolution for different orientations of the magnetic field is shown in Fig. \ref{fig:rotation-scans} and discussed below. 

\begin{itemize}

    \item {\bf $y-z$ plane:} in Fig. \ref{fig:rotation-scans}(a) we show the evolution of $\Delta R_{zz}/R_{zz}B$ when the magnetic field ${\bf B}$ is confined to the $y-z$ plane. In this case, the relevant component of the polarization is the large ${\bf P}\parallel\hat{\bf x}$ component. As ${\bf B}$ moves away from the $\hat{\bf y}-$axis by an angle $\theta$ and towards the $\hat{\bf z}-$axis, the direction of the vector product ${\bf P}\times{\bf B}$ moves away from the $\hat{\bf z}-$axis by the same angle $\theta$. Consequently, the evolution of the normalized resistance is determined by
    \begin{equation}
        {\bf I}\cdot({\bf P}\times{\bf B})\propto\cos(\theta),
    \end{equation}
    where $\theta$ is the angle between ${\bf B}$ and the $\hat{\bf y}-$axis, as shown in the inset of Fig. \ref{fig:rotation-scans}(a).

    \item {\bf $x-y$ plane:} in Fig. \ref{fig:rotation-scans}(b) we show the evolution of $\Delta R_{zz}/R_{zz}B$ when the magnetic field ${\bf B}$ is confined to the $x-y$ plane. Here, the relevant component of the polarization is again the large ${\bf P}\parallel\hat{\bf x}$ component. However, as ${\bf B}$ moves away from the $\hat{\bf y}-$axis by an angle $\phi$ and towards the $\hat{\bf x}-$axis, the direction of the vector product ${\bf P}\times{\bf B}$ is remains fixed along the $\hat{\bf z}-$axis, and only its modulus decreases. The evolution of the normalized resistance is determined by
    \begin{equation}
        {\bf I}\cdot({\bf P}\times{\bf B})\propto
        \sin\left(\frac{\pi}{2}-\phi\right)\propto\cos(\phi),
    \end{equation}
    where $\phi$ is the angle between ${\bf B}$ and the $\hat{\bf y}-$axis, as shown in the inset of Fig. \ref{fig:rotation-scans}(b).

    \item {\bf $x-z$ plane:} in Fig. \ref{fig:rotation-scans}(c) we show the evolution of $\Delta R_{zz}/R_{zz}B$ when the magnetic field ${\bf B}$ is confined to the $x-z$ plane. In this case, the relevant component of the polarization is the small ${\bf P}\parallel-\hat{\bf y}$ component. As ${\bf B}$ moves away from the $\hat{\bf z}-$axis by an angle $\alpha$ and towards the $\hat{\bf x}-$axis, the direction of the vector product ${\bf P}\times{\bf B}$ also moves away from the $-\hat{\bf x}-$axis by the same angle $\alpha$. The evolution of the normalized resistance is determined by
    \begin{equation}
        {\bf I}\cdot({\bf P}\times{\bf B})\propto
        \cos\left(\frac{\pi}{2}-\alpha\right)\propto -\sin(\alpha),
    \end{equation}
    where $\alpha$ is the angle between ${\bf B}$ and the $\hat{\bf z}-$axis, as shown in the inset of Fig. \ref{fig:rotation-scans}(c). 

\end{itemize}

We point out that the scale of $\Delta R_{zz}/R_{zz}B$ in Fig. \ref{fig:rotation-scans}(c) is two orders of magnitude smaller than in Figs. \ref{fig:rotation-scans}(a) and (b). This difference arises because, in Fig. \ref{fig:rotation-scans}(c), the relevant component for the polarization is the ${\bf P}\parallel-\hat{\bf y}$, which is $20$ times smaller than the ${\bf P}\parallel\hat{\bf x}$ component. Furthermore, the data shown in Fig. \ref{fig:rotation-scans}(c) was collected at a higher temperature ($T=2$K) and voltage ($V=20$V), both of which contribute to a reduction of the proportionality coefficient $\gamma^\pm(T,V)$. 

\section{Numerical Method}

Density Functional Theory (DFT) was employed to calculate the macroscopic polarization of two-dimensional Te systems using the Quantum Espresso package \cite{Giannozi2009}. We utilized norm-conserving Rappe-Rabe-Kaxiras-Joannopoulos (RRKJ) pseudopotentials and the Perdew-Burke-Ernzerhof (PBE) exchange-correlation functional. Initially, we relaxed both the cell dimensions and atomic positions for an orthorhombic crystal structure of the two-layer $\alpha-\text{Te}$ phase, which contains six Te atoms per unit cell. We then relaxed the atomic positions for the $\alpha'-\text{Te}$ phase, using the same geometry parameters for both phases.

The optimized lattice vectors were determined as follows: $a=4.39$ \AA\ (parallel to the surface),  $b=5.89$ \AA\ (coil direction), and $c=19.76$ \AA\ (perpendicular to the surface),  yielding a vacuum spacing of approximately $14$ \AA\ between layers.  This separation ensures the system can be treated as isolated and periodic in two dimensions \cite{Sohier_2017}.  The interatomic distances within the coil direction were found to be $2.93$ \AA\ and $2.89$ \AA\ in relation to the central atom in the unit cell coil.  

Next, we performed self-consistent calculations under an applied electric field between $\pm 0.002$  Ry a.u.  ($=36.3609\times 10^{10}$ V/m) along the $a-\text{axis}$. The calculations used a plane-wave cutoff of $60$ Ry and a $12 \times 8 \times 1$ Monkhorst-Pack k-point grid.  The Berry phase was then computed along the three reciprocal lattice vectors using a 24 k-points grid.

For the charge density calculations, we employed the SIESTA code \cite{Soler_2002} with the van der Waals (VDW-DRSLL) exchange-correlation functional and a double-zeta polarization basis set for several two-dimensional Te structures with different number of layers. We saw that the charge polarization decreases with the number of layers, being more prominent in two layers Te system. The calculations were performed using a $12 \times 1 \times 8$ Monkhorst-Pack k-point grid with $300$ Ry plane-wave cutoff.  The unit cell had the following lattice vectors: $a=4.57$ \AA\, $b=27.92$ \AA\ and $c=6.13$ \AA\ (note that the orientations of the $b$ and $c$ lattice vectors differ from those in the Quantum Espresso calculations).  The structures were relaxed until the forces between atoms were less than $10^{-3}$ eV/\AA .  Finally, the macroscopic polarization components were computed using the geometric Berry phase method, with a k-point grid of 10 points along the polarization direction and a $3\times 3$ grid perpendicular to the surface.


\bibliography{biblio}
